\documentclass[times,preprint,3p]{elsarticle}

\journal{Journal of Computational Physics}

\usepackage{multirow}

\usepackage{hyperref}
\usepackage{color}
\usepackage{amsfonts}
\usepackage{amsmath}
\usepackage{graphicx}
\usepackage{rotating}
\usepackage{mathtools}
\usepackage[shortlabels]{enumitem}

\usepackage{siunitx} 
\usepackage{graphicx}
\usepackage{subcaption}

\usepackage{booktabs}

\newcommand{\bff}{\mathbf{f}}

\newcommand{\bfx}{\mathbf{x}}
\newcommand{\bfy}{\mathbf{y}}

\newcommand{\half}{\frac{1}{2}}
\newcommand{\pd}[2]{\frac{\partial #1}{\partial #2}}
\newcommand{\pdd}[2]{\frac{\partial^{2} #1}{\partial #2^{2}}}

\newcommand{\mm}[1]{\rm mm}

\newcommand{\beq}{\begin{equation}}
\newcommand{\eeq}{\end{equation}}
\newcommand{\bea}{\begin{eqnarray}}
\newcommand{\eea}{\end{eqnarray}}

\newcommand{\dt}{\Delta t}
\newcommand{\dx}{\Delta x}
\newcommand{\dy}{\Delta y}

\newcommand{\bit}{\begin{itemize}}
\newcommand{\eit}{\end{itemize}}
\newcommand{\ben}{\begin{enumerate}}
\newcommand{\een}{\end{enumerate}}

\newcommand{\bF}{\mathbf{F}}
\newcommand{\bG}{\mathbf{G}}

\newcommand{\bL}{\mathbf{L}}

\newcommand{\bR}{\mathbf{R}}

\newcommand{\bU}{\mathbf{U}}
\newcommand{\bV}{\mathbf{V}}
\newcommand{\bW}{\mathbf{W}}

\newcommand{\bg}{\mathbf{g}}

\newcommand{\bx}{\mathbf{x}}

\DeclareMathOperator*{\argmin}{arg\,min}

\makeatletter
\def\@xfootnote[#1]{%
  \protected@xdef\@thefnmark{#1}%
  \@footnotemark\@footnotetext}
\makeatother

\usepackage{url}
\usepackage{xcolor}
\definecolor{newcolor}{rgb}{.8,.349,.1}

\begin{document}
\begin{frontmatter}

\title{A single-step third-order temporal discretization with
Jacobian-free and Hessian-free formulations for finite difference methods}

\author[1]{Youngjun {Lee}}
\ead{ylee109@ucsc.edu}
\author[1]{Dongwook {Lee}\corref{cor1}}
\cortext[cor1]{Corresponding author: Tel.: +1-831-502-7708}
\ead{dlee79@ucsc.edu}

\address[1]{Department of Applied Mathematics, The University of California, Santa Cruz, CA, United States}

\begin{abstract}
Discrete updates of numerical partial differential equations (PDEs)
rely on two branches of temporal integration.
The first branch is the widely-adopted, traditionally popular
approach of the method-of-lines (MOL) formulation,
in which multi-stage Runge-Kutta (RK) methods have shown
great success in solving ordinary differential equations (ODEs)
at high-order accuracy. The clear separation between the temporal
and the spatial discretizations of the governing PDEs makes the
RK methods highly adaptable. In contrast, the second branch of
formulation using the so-called Lax-Wendroff procedure
escalates the use of tight couplings between the spatial
and temporal derivatives to construct high-order approximations
of temporal advancements in the Taylor series expansions.
In the last two decades, modern numerical methods have
explored the second route extensively and have proposed
a set of computationally efficient single-stage, single-step
high-order accurate algorithms. In this paper, we present
an algorithmic extension of the method called the
Picard integration formulation (PIF) that belongs to the
second branch of the temporal updates.
The extension presented in this paper furnishes ease of
calculating the Jacobian and Hessian terms necessary
for third-order accuracy in time.
\end{abstract}

\end{frontmatter}
%
%
%
%

\section{Introduction}\label{sec:introduction}
The evolution of high-performance computing (HPC) drives
the advent of high-order discrete methods
in computational fluid dynamics (CFD) research area.
As the hardware progression of the memory capacity per compute core
has become gradually saturated~\cite{attig2011trends, subcommittee2014top},
the HPC community has been compelling to find more efficient ways
that can best exercise computing resources in pursuing computer simulations.
Developing highly efficient numerical algorithms has
become an important subject in CFD research fields.
Modern practitioners have relentlessly delved into advancing
high-arithmetic-intensity models that can increase numerical accuracy per
degree of freedom, while at the same time, can operate with reduced
memory requirements and data transfers in HPC architectures.
One such computing paradigm is to promote high-order methods
in which high arithmetic intensity is achieved by putting to use an increasing number of
higher order terms.
High-order methods (third-order or higher) avail themselves of
an improved figure-of-merit in terms of
reaching a target numerical error per computational time faster
than the other counterpart, i.e., lower-order methods.

Within the broad framework of finite difference method (FDM) and
finite volume method (FVM) discretization schemes,
it has proven that high-order data interpolations and reconstructions
play a key role to reduce numerical errors
for solving discrete partial differential equations (PDEs)
on a given grid resolution.
As seen in the success of the
Piecewise Parabolic Method (PPM)~\cite{colella1984piecewise} and
Weighted Essentially Non-Oscillatory (WENO) method~\cite{jiang1996efficient},
to name a few, high-order interpolation/reconstruction schemes
produce more accurate numerical solutions
with a faster rate of convergence-to-solution on a limited computing resource.
It is of great advantage to adopting the increased intensity of floating-point operations
in scientific simulations on modern HPC architecture,
because otherwise increasing grid resolutions to reach the desired solution convergence
will require a larger memory footprint
which is typically bounded for most HPC systems.

To acquire highly accurate numerical solutions,
a high-order temporal discretization
scheme ought to be considered alongside
the high-order spatial data interpolation/reconstruction.
In the context of FDM and FVM formulations,
numerical errors arise from both spatial and temporal axes
since the solution lies on the spatio-temporal plane.
In most compressible simulations of hyperbolic PDE systems with strong shocks,
the magnitude of the maximum wave speed could most likely surpass the size of the
smallest grid delta. 
Thus, the timestep size $\Delta t$ is typically
smaller than the minimum size of grid deltas,
$\min_{d=x,y,z}\{\Delta d\}$,
subject to the hyperbolic CFL stability requirement.
Under this condition, it is seen that
the temporal error of an $m$-th order time integrator is
smaller than the spatial error of an $n$-th order spatial solver,
i.e.,
$\mathcal{O}(\Delta t^m) < \mathcal{O}(\sum_{d=x,y,z} \Delta d^n)$ in 3D,
with the usual assumption of $m \le n$.
This, however, flips to the other end in which
the temporal error becomes more dominant over the spatial error
as the computational need necessitates a finer grid resolution
to resolve smaller physical length scales.

For decades, the strong stability preserving Runge-Kutta (SSP-RK)
time integrator~\cite{gottlieb1998total, gottlieb2001strong, gottlieb2011strong}
has been considered to be the \textit{standard} strategy
for an extensive range of high-order numerical schemes for PDE solvers.
The SSP-RK schemes have proven its high fidelity which guarantees
not only high-order accuracy but also numerical
stability with total variation diminishing (TVD) property.
The chief objective of SSP-RK schemes is to maintain the strong stability property (SSP)
at high-order accuracy by sequentially applying convex combinations of
the first-order forward Euler method as a building-block
at each sub-stage~\cite{gottlieb2001strong}.
The desired total variation diminishing (TVD) property
is achieved if each of the sub-stage is TVD\@. The Runge-Kutta schemes
pertaining this TVD property are known as SSP-RK schemes.
Designed this way, the SSP-RK procedure requires $s$ multiple
sub-stages to advance the solution by
a single time step $\Delta t$ at $m$-th accuracy.

Of the most popular choice among various SSP-RK solvers
is the three-stage, third-order RK3, given that, in practice,
spatial accuracy is often considered to 
carry more weights
than temporal accuracy in designing 
higher accurate spatial models~\cite{balsara2000monotonicity,mignone2010high}.
In contrast, devising a fourth-order SSP-RK4 is more involved to meet the favorable
SSP property. Several theoretical studies have shown that
a fourth-order SSP-RK4 cannot be formulated with just
four sub-stages~\cite{gottlieb1998total}, meaning that
the classical four-stage, fourth-order RK4 is not SSP\@.
Other authors have demonstrated that SSP-RK4 with positive coefficients
could be constructed with an increasing number of sub-stages from five
up to eight~\cite{spiteri2002new,spiteri2003non}.
We remark that, to provide stable, non-oscillatory predictions
in the presence of shocks, it is safer to employ SSP-RK schemes than
to use non SSP-RK schemes.
However, such SSP-RK schemes experience theoretical barriers in attaining
the maximum available order and the maximum
CFL coefficient~\cite{ruuth2002two,giuliani2019optimal}.
Besides, the spatial interpolation/reconstruction, as well as boundary conditions,
should be applied at each sub-stage, which makes the overall procedure of SSP-RK
computationally expensive. In parallel simulations,
these operations also increase the footprint
of data movements between node communications
as the number of sub-stages grows.
This very nature of SSP-RK makes it less efficient for
massively parallel simulations particularly when
the level of adaptive mesh refinement (AMR) progressively builds up
around interesting features in simulations.

As a means of circumventing the said issues in SSP-RK, practitioners have
taken a different route of providing a high-order temporal updating strategy for solving
numerical PDEs. The core design principle lies in formulating
a conservative temporal integrator
which works for nonlinear PDEs in multiple spatial dimensions
with the equivalent high-order accuracy as in SSP-RK,
but, this time, in a single-stage, single-step update.
The effort in this direction has resulted in what is now widely known as
Arbitrary high order DERivative (ADER) method,
which was first introduced in~\cite{toro2001towards} for linear equations.
Since then, ADER schemes have gone through several generations of breakthrough
by numerous authors with the common goal of meeting
the high-order requirement in a compact single-step update.

In the developments lead by Toro and his collaborators,
the single-step, high-order ADER
solution update is attained by solving a series of
generalized (or high-order) Riemann problems
(GRPs)~\cite{van1979towards,ben1984second,ben2007hyperbolic}
to compute the coefficients corresponding to
the high-order terms in the power series expansion
of each conservative variable in
time~\cite{titarev2002ader, titarev2005ader, toro2001towards}.
Solving GRPs involves two steps.
The first step is to solve one set of
classical nonlinear Riemann problems for the leading term.
The second step is to solve another set of classical~--~but linearized~--~Riemann problems
for the rest higher derivative terms via the Cauchy-Kowalewski procedure, which
exercises a sequence of full couplings between spatial and temporal derivatives.

The ADER formulation has been further taken to a more modern direction.
Dumbser \textit{et al.}~\cite{dumbser2008unified} extended the original ADER scheme
with multiple quadrature points to an efficient quadrature-free ADER scheme
with a space-time averaged numerical flux, and the authors
applied the scheme to integrate a discontinuous Galerkin (DG) solver.
Balsara \textit{et al.}~\cite{balsara2009efficient} presented a new compact ADER framework
that replaces the usual Cauchy-Kowalewski procedure in the original ADER formalism
with a local continuous space-time
Galerkin formulation up to fourth-order, and called the new approach
ADER-CG (CG for continuous Galerkin).
A comprehensive review on the recent developments of
ADER-CG has been reported in~\cite{balsara2017higher}.
In~\cite{balsara2013efficient}, the ADER-CG schemes are shown to be approximately
twice faster than the RK methods at the same order of accuracy.
It is also shown that ADER is highly adaptable to the
AMR grid configuration~\cite{dumbser2013ader}.
The ADER formulation has been further extended to compressible
dissipative flows in both hydrodynamics and magnetohydrodynamics
in the ADER-DG framework~\cite{fambri2017space}
using the
MOOD method~\cite{clain2011high,diot2012improved,
diot2013multidimensional,boscheri2015direct,loubere2014new}
as a posteriori detection to
enhance the shock-capturing capability;
a comparison study of the space-time predictor calculation
based on primitive, conservative, and characteristic
variables~\cite{zanotti2016efficient};
a new approach of using the so-called automatic differentiation
(or the differential transform method) to reduce the computational
cost of ADER schemes called the ADER-Taylor method~\cite{norman2012multi, norman2013algorithmic, norman2014weno}, etc.

Unlike the broad usage of SSP-RK in various discrete PDE solvers,
the developments of ADER mentioned above have been exclusively
applied to FV and DG methods, but FD methods.
This is mainly because the fundamental principle of obtaining high-order accuracy
in the original ADER scheme relies on solving generalized (or high-order) Riemann
problems, which are the characteristic building blocks of FV and DG methods.
In this regard, perhaps there could be a pathway to devise an ADER scheme for FDM
within a context of a new FDM formulation called
FD-Prim~\cite{del2007echo,chen2016fifth,reyes2019variable}
in which solving Riemann problems is fully utilized as a new way of forming
high-order FD fluxes.

Categorically, ODE solvers can be classified as members of
multi-stage multi-derivative
methods~\cite{hairer1973multistep,seal2014high}.
For example, RK schemes are of the multi-stage type as they
attain high-order accuracy
via utilizing multi-stages as a primary mechanism,
whereas ADER methods are of the multi-derivative type
since they explore multi-derivatives as a central
engine.
In this paper, we are interested in constructing~--~for general FD solvers~--
a single-stage, single-step temporal updating algorithm
that belongs to the class of multi-derivative methods.
We follow the recent approach of
the Picard integral formulation (PIF) method~\cite{christlieb2015picard, seal2016explicit}
for achieving high-order temporal accuracy in FDM, where
the authors designed a high-order temporal discretization by constructing
a high-order approximation to the \textit{time-averaged fluxes} over $[t^n, t^{n+1}]$.
The PIF discretization introduced in~\cite{christlieb2015picard}
attains third-order temporally accurate numerical fluxes by computing
the coefficients of the Taylor expansion of the averaged fluxes up to third-order
via the Lax-Wendroff (or Cauchy–Kowalewski) procedure which converts
the high-order temporal derivatives terms in the Taylor expansion
to spatial derivative terms.

As the resulting PIF scheme incorporates the needed high-order discretization
directly into the numerical fluxes, the governing hyperbolic conservation law,
$\partial_t\bU +\nabla\cdot \mathcal{F}(\bU) = 0$,
can be inherently satisfied
at least on a simple uniform Cartesian geometry.
In contrast, 
in the one-step temporal discretization
methods reported previously in~\cite{qiu2003finite, jiang2013alternative} for FDM,
which are analogously based on the Lax-Wendroff procedure,
the high-order expansion was performed on conserved variables
with the Lax-Wendroff conversion of temporal derivatives to spatial derivatives.
The mechanism that made these approaches conservative
is set by discretizing those spatial derivatives via central differencing,
which is less intrinsic than that of PIF in terms of constructing
a conservative update in a discrete sense.
Accordingly, in PIF, there are a few more benefits of working on the time-averaged fluxes.
As shown in~\cite{seal2016explicit}, an extra manipulation of numerical fluxes
(e.g., a hybrid numerical flux as a linear combination of a low- and high-order flux
to preserve positivity) does not break the hyperbolic conservation law.
In addition, as will be seen throughout the present study,
the high-order handling of the time-averaged fluxes can serve
as an extra standalone step, apart from the rest of the conventional FDM discretization steps
such as the flux splitting, high-order flux reconstruction, as well as the final temporal update.
As a consequence, the PIF step can readily replace any existing SSP-RK updating step,
boosting the overall performance gain by a factor of two.

The primary objective of the current work lies in providing increased flexibility
and ease of computing the needed flux Jacobian and Hessian tensor terms
in the original third-order PIF method~\cite{christlieb2015picard, seal2016explicit}.
To do that, we offer an extended algorithmic development that permits
the equivalent single-step, third-order PIF method that does not involve
the analytic evaluations 
of the Jacobian and Hessian terms as described
in~\cite{christlieb2015picard, seal2016explicit}.
We call our new method  a system-free PIF method (or SF-PIF in short)
to reflect the added capability of the proposed Jacobian-free and Hessian-free approach.

We organize the paper as follows. In Section~\ref{sec:finite-diff-meth}, we give an overview
of the general principles in the PIF-based FD discretization.
The new concept of our system-free approach to approximating
the flux Jacobian and Hessian tensor terms is introduced in Section~\ref{sec:system-free}.
We test the new SF-PIF method on a series of 1D and 2D benchmark problems
in the system of the Euler equations as well as the shallow water equations.
The test results are presented in Section~\ref{sec:results}.
We conclude our paper with a brief summary in Section~\ref{sec:conclusion}.


\section{PIF Method}\label{sec:finite-diff-meth}
We are interested in solving the general conservation laws of hyperbolic PDEs
in 1D and 2D, predicting numerical solutions with the target third-order accuracy in time.
To present a mathematical foundation of our method, we begin with
the following general system of equations in 2D,
\begin{equation}\label{eq:gov}
    0= \pd{\bU}{t} + \nabla \cdot \mathcal{F}(\bU) = \pd{\bU}{t} + \pd{\bF(\bU)}{x} + \pd{\bG(\bU)}{y}.
\end{equation}

Denoted by \( \bU \) is a vector of conservative variables, \( \bF \) and \( \bG \)
are the flux functions in $x$- and $y$- directions, respectively.
Here, we apply the Picard integral formulation (PIF)~\cite{christlieb2015picard}
by taking a time average of Eq.~\eqref{eq:gov} within a single timestep \( \dt \) over an interval
$[t^n, t^n+\dt] = [t^n, t^{n+1}]$,
\begin{equation}\label{eq:pif}
    \bU^{n + 1} = \bU^{n} - \dt \left( \pd{\bF^{avg}}{x} + \pd{\bG^{avg}}{y} \right),
\end{equation}
where \( \bF^{avg} \) and \( \bG^{avg} \) represent the time-averaged fluxes
in \( x \)- and \( y \)-direction respectively,
\begin{equation}\label{eq:avg-flx}
    \bF^{avg} (\bx) \equiv \frac{1}{\dt} \int^{t^{n + 1}}_{t^{n}} \bF(\bU(\bx, t)) \mathop{dt}, \quad\quad
    \bG^{avg} (\bx) \equiv \frac{1}{\dt} \int^{t^{n + 1}}_{t^{n}} \bG(\bU (\bx, t)) \mathop{dt}, \quad\quad
    \bx = (x, y) \in \mathbb{R}^2.
\end{equation}

In order to get a spatially discretized solution \( \bU_{i,j} \equiv \bU(\bx_{ij}) \), we wish to express
Eq.~\eqref{eq:pif} with numerical fluxes \( \hat{\bff} \) and \( \hat{\bg} \) at cell interfaces defined as,
\begin{equation}\label{eq:num-flx}
    \left. \pd{\bF^{avg}}{x} \right|_{\bx = \bx_{ij}} =
    \frac{1}{\dx} \left( \hat{\bff}_{i + \frac{1}{2}, j} -
                                \hat{\bff}_{i - \frac{1}{2}, j} \right) +
                                \mathcal{O} (\dx^{p} + \dt^{q}), \quad \quad
    \bx_{ij} = (x_i, y_j).
\end{equation}
The numerical flux in \( y \)-direction, \( \hat{\bg} \), is defined in a similar fashion.
Now we are able to express Eq.~\eqref{eq:gov} in a fully discretized form as,
\begin{equation}\label{eq:discrete-soln}
    \bU^{n + 1}_{i,j} = \bU^{n}_{i, j} -
    \frac{\dt}{\dx} \left( \hat{\bff}_{i + \frac{1}{2}, j} -
                                 \hat{\bff}_{i - \frac{1}{2}, j} \right) -
    \frac{\dt}{\dy} \left( \hat{\bg}_{i, j + \frac{1}{2}} -
                                  \hat{\bg}_{i, j - \frac{1}{2}}   \right).
\end{equation}
We remark that the derived governing form in
Eq.~\eqref{eq:discrete-soln} for PIF is something
in between those of FVM and FDM. It is different from
that of FVM that it does not carry any spatial average but the temporal average.
It is also different from that of FDM
that it does involve the temporal average in the fluxes
in Eq.~\eqref{eq:avg-flx} to which the numerical
fluxes \( \hat{\bff} \) and \( \hat{\bg} \) approximate.

We aim to get a high-order approximated solution for \( \hat{\bff} \) and \( \hat{\bg} \),
\textit{both in space and time}, by which the update in Eq.~\eqref{eq:discrete-soln}
enables us to update the solution \( \bU^{n+1}_{i,j} \) in a single step
at the desired $p$-th and $q$-th accuracy in space and time.
To do this, we first follow the standard convention in FDM
in which we treat the pointwise $x$-flux $\bF(x,y_j)$ 
as a 1D cell average of
an auxiliary function $\hat{\bF}$ in 1D,
\begin{equation}\label{eq:aux_eqn}
    \bF(x,y_j) = \frac{1}{\Delta x} \int_{x  -{\frac{\dx}{2}}}^{x + {\frac{\dx}{2}}} \hat{\bF}(\xi, y_j) d\xi.
\end{equation}
Then the analytic derivative of $\bF$ at $x=x_{i}$ in $x$-direction becomes
\begin{equation}\label{eq:aux_analytic_deriv}
    \left. \frac{\partial \bF}{\partial x}\right|_{x=x_{i}} =
    \frac{1}{\dx} \left(\hat{\bF}(x_{i+\frac{1}{2}}, y_j) - \hat{\bF}(x_{i-\frac{1}{2}}, y_j) \right).
\end{equation}
Comparing Eq.~\eqref{eq:aux_analytic_deriv} and Eq.~\eqref{eq:num-flx},
our goal is to be achieved if we can define the numerical flux $\hat {\bff}$ that satisfies
the following relationship with $\hat{\bF}$,
\begin{equation}\label{eq:num-flx-approx}
    \hat{\bff}_{i \pm \frac{1}{2}, j} =
    \hat{\bF} (x_{i \pm \frac{1}{2}}, y_j) + \mathcal{O} (\dx^{p} + \dt^{q}).
\end{equation}
Mathematically speaking, the inverse problem of Eq.~\eqref{eq:aux_eqn} is exactly the same as
the conventional 1D reconstruction problem in FVM, the operation of which is specifically designed to
find the primitive function value $\hat{\bF}$ at a certain location (mostly, $x_{i\pm 1/2}$) in the $i$-th cell,
given the integral-averaged (or volume-averaged) values ${\bF}$ at nearby stencil points
as input. Namely, this can be written as
\begin{equation}\label{eq:inverse_reconst}
    \hat{\bF}(\xi, y_j) = \mathcal{R}\left(\bF_{i-r, j}, \dots, \bF_{i+r, j}\right) + \mathcal{O} (\dx^p),
    \quad
    \xi \in [x_{i-\frac{1}{2}}, x_{i+\frac{1}{2}}],
\end{equation}
where $\mathcal{R}$ is a $p$-th order accurate reconstruction operator that takes inputs
of the pointwise fluxes $\bF$ at the $x$ neighboring stencil $[x_{i-r,j}, \dots, x_{i+r,j}]$ with fixed $y=y_j$.

We can split the task in Eq.~\eqref{eq:num-flx-approx} and Eq.~\eqref{eq:inverse_reconst}
in two different stages in the following sequential order.
The first step is to attain $q$-th order temporal accuracy,
followed by the second step that provides $p$-th order spatial accuracy:
\begin{enumerate}
    \item[] \textbf{Step 1}
     Approximation of the time-averaged flux~--
     Obtain a $q$-th order
     approximated solution $\bF^{appx}(\bx)$ to Eq.~\eqref{eq:avg-flx}:
        \begin{equation}\label{eq:approx-avg-flx}
        \bF^{avg} (\bx) \equiv \frac{1}{\dt} \int^{t^{n + 1}}_{t^{n}} \bF(\bU(\bx, t)) \mathop{dt}
            = \bF^{appx} (\bx)+ \mathcal{O}(\dt^{q}).
        \end{equation}
    \item[] \textbf{Step 2}
    Approximation of the high-order spatial reconstruction~--
    Use a traditional high-order finite difference method with the
    \textit{approximated time-averaged} fluxes $\bF^{appx}_{i,j}$ from \textbf{Step 1},
    instead of the conventional choice of pointwise fluxes, i.e., $\bF_{i,j}$ at $t=t^n$:
        \begin{equation}\label{eq:recons}
            \hat{\bff}_{i + \frac{1}{2}, j} =
            \mathcal{R}\left(\mathcal{FS}\left(\bF^{appx}_{i-r, j}, \dots,
                \bF^{appx}_{i+r+1, j} \right) \right)
            + \mathcal{O}(\dx^{p}).
        \end{equation}
    Here, \( \mathcal{FS}(\cdot) \) represents
    a characteristic flux splitting in the positive and negative directions, and
    \( \mathcal{R}(\cdot) \) is a high-order reconstruction scheme which is also used for
    FVM formulations.
\end{enumerate}
In this study, we adopt the fifth-order WENO method (or WENO5)~\cite{jiang1996efficient}
for the reconstruction in \textbf{Step 2},
combined with the Rusanov Lax–Friedrichs
flux splitting described in~\cite{mignone2010high}, in order to
attain high-order accuracy in space domain.

Before we discuss \textbf{Step 1} in details,
we give a brief description on the flux splitting \( \mathcal{FS} (\cdot) \), given
the fact that we perform the splitting on the time-averaged
approximated fluxes,
$\bF^{appx}_{\ell,j}$, $i-r \le \ell \le i+r+1$,
whose temporal state is not at $t=t^n$
but at an average state over $[t^n, t^{n+1}]$.
This is different from the conventional FDM splitting which performs
on the pointwise fluxes $\bF$ at $t=t^n$.
Following~\cite{mignone2010high}, the Rusanov Lax-Friedrichs splitting
projects the temporally averaged fluxes
$\bF^{appx}_{\ell,j}$ from \textbf{Step 1}
to the left- and the right-going parts
according to the characteristic decomposition
of the Jacobian matrix,
\begin{equation}\label{eq:Jacobian}
    \left. \pd{\bF}{\bU}\right|_{\bU^{n}_{i+\frac{1}{2},j}}=
    \bR_{i+\frac{1}{2},j} \Lambda_{i+\frac{1}{2},j} \bL_{i+\frac{1}{2},j}, \quad
    \bU^{n}_{i+\frac{1}{2},j} = \frac{\bU^{n}_{i,j} + \bU^{n}_{i+1,j}}{2},
\end{equation}
where $\bR$ and $\bL$ are the corresponding matrices of right and left eigenvectors,
and $\Lambda$ is the diagonal matrix whose diagonal entries are eigenvalues.
%
%
The projection proceeds to construct $s$ different
left-going (\( - \)) and right-going (\( + \))
\textit{characteristic states of the averaged-fluxes},
denoted as $\bV^{k,\pm}_{(i+\half):{s},j}$,
to the cell interface $(i+\half,j)$ as
\begin{equation}\label{eq:flux-splitting}
    \bV^{k,+}_{(i+\half):{s},j} = \mathcal{FS}^{+}(\bF^{appx}_{s,j})
    =\half \bL_{i+\half,j}^k \cdot \left(\bF^{appx}_{s,j} + \alpha^k \bU_{s,j} \right), \quad
    \bV^{k,-}_{(i+\half):{s},j} = \mathcal{FS}^{-}(\bF^{appx}_{s',j})
    =\half \bL_{i+\half,j}^k \cdot \left(\bF^{appx}_{s',j} - \alpha^k \bU_{s',j} \right),
\end{equation}
where the sub-index $s$ ranges from $i-r, \dots, i+r$,
while at the same time, $s'=2i - s + 1$.
The super-index $k$ represents each characteristic field. 
The coefficient $\alpha^k$ is chosen to be the maximum absolute value of the $k$-th characteristic
speed over the entire computational domain, resulting in the so-called \textit{global} Lax-Friedrichs flux splitting.
Here, we point out that, in conventional FDM, the
temporal states of the flux and conservative vectors,
the left eigenvectors, and $\alpha^k$ in Eq.~\eqref{eq:flux-splitting}
are all consistently at $t=t^n$.
For PIF, though, the flux vector $\bF^{appx}_{s,j}$ is at the averaged temporal state, while
the others are at $t=t^n$. Ideally, one can also choose to compute a similar temporally averaged
conservative vectors $\bU^{appx}_{s,j}$, use them to compute $\bL_{i+\half,j}^k$ and $\alpha^k$,
in order to provide a consistent temporal handling across all the quantities in Eq.~\eqref{eq:flux-splitting}.
However, we have found that the level of numerical stability as well as the solution accuracy (e.g.,
preserving the expected symmetries in some symmetry-preserving problems) tend to become
less maintained in that approach. This observation has made us to decide that our flux splitting
uses theoretically inconsistent temporal states between $\bF^{appx}_{s,j}$ and the other quantities.
As will be shown in Section~\ref{sec:results}, this treatment does not affect the overall solution accuracy
and stability. The same choice has been also made in the original PIF method~\cite{christlieb2015picard}.
The rest of the flux splitting procedure follows to compute the numerical fluxes $\hat{\bff}_{i\pm \half,j}$
via a high-order FDM reconstruction operator, $\mathcal{R}(\cdot)$. More specifically, we write,
\begin{equation}\label{eq:flux-splitting-final}
    \hat{\bff}_{i+\half,j} = 
    \sum\limits_{k}\left( \hat{\bV}_{i+\half,j}^{k,+} + \hat{\bV}_{i+\half,j}^{k,-}\right) \bR^{k}_{i+\half,j},
\end{equation}
which is a linear combination (over the $k$ characteristic fields) of
the coefficients $\hat{\bV}_{i+\half,j}^{k,\pm}$ obtained by
solving a high-order reconstruction,
\begin{equation}\label{eq:weno-recons}
    \hat{\bV}_{i+\half,j}^{k,\pm} =
    \mathcal{R}\left(
        \bV^{k,\pm}_{(i+\half):{s},j}
    \right)
    =
    \mathcal{R}\left(
        \mathcal{FS}^{+}(\bF^{appx}_{s,j}),
        \mathcal{FS}^{-}(\bF^{appx}_{s',j})
    \right),
    \quad s=i-r, \dots, i+r,
    \quad \mbox{with} \;\;\;
    s'=2i-s+1.
\end{equation}
In this paper we use $r=2$ for the fifth-order WENO5 method~\cite{jiang1996efficient}.
We assume that readers are familiar with the WENO5 procedure in Eq.~\eqref{eq:weno-recons};
hence it is omitted in this paper. Interested readers are
encouraged to refer to~\cite{jiang1996efficient,mignone2010high}.

Let us now focus on \textbf{Step 1}, our main interest in this study, which comprises
achieving a high-order temporal accuracy
of the numerical fluxes $\bF^{appx}_{i,j}$.
Traditionally, this is accomplished
by a multi-stage time integrator, such as an $s$-stage, $m$-th order RK method.
However, as briefly discussed in Section~\ref{sec:introduction},
a multi-stage method is computationally expensive,
and in practice, is not suitable on an adaptive mesh refinement (AMR) grid configuration
due to the increasing amount of calculation and data movements
as the refinement level grows in simulations.
On the other hand, by being a single-stage, single-step method,
a predictor-corrector-type of time integration schemes that are based on
the Lax-Wendroff (or Cauchy-Kowalewski) procedure has an advantage
to reduce a plenty of computational costs and is favorable to be
implemented on an AMR grid configuration
while maintaining a high-order accuracy.

The PIF method is to be considered as an approach
that belongs to this time integration category, featuring a new formulation of involving
the concept of time-averaged fluxes.
Using a time-averaged quantity in a predictor step has been widely used in FVM formulations,
including the original Godunov scheme~\cite{godunov1959difference},
piecewise-type methods~\cite{colella1984piecewise, lee2017piecewise}
and arbitrary derivative (ADER)
schemes~\cite{titarev2002ader, titarev2005ader, balsara2009efficient}.
Recently, \textit{Seal et al.}~\cite{christlieb2015picard, seal2016explicit} extended this approach
to the FDM formulation, by way of introducing a Picard integration
as described in Eq.~\eqref{eq:pif}, in which the time-averaged fluxes in~Eq.~\eqref{eq:avg-flx}
are approximated through Taylor expansion of the integrand around $t^{n}$.
For example, in the third-order temporal PIF method, the time-averaged flux
$\bF^{avg}(\bx)$ is approximated by
an integral of the Taylor expansion of the pointwise flux $\bF(\bx,t)$ around $t^n$,
\begin{equation}\label{eq:taylor3}
    \begin{split}
        \bF^{avg}(\bx)
        &= \frac{1}{\dt} \int^{t^{n + 1}}_{t^n} \bF(\bx, t) \mathop{dt} \\
        & = \frac{1}{\dt} \int^{t^{n + 1}}_{t^n}
               \left[ \bF(\bx,t^n) +
               (t-t^n) \left.\pd{\bF(\bx,t)}{t}\right|_{t = t^{n}} +
               \frac{(t-t^n)^2}{2!}\left.\pdd{\bF(\bx,t)}{t}
               \right|_{t = t^{n}} + \cdots
               \right]\mathop{dt} \\
        & = \bF(\bx,t^n) + \frac{\dt}{2!} \left.\pd{\bF(\bx,t)}{t}\right|_{t = t^{n}}
                           + \left.\frac{\dt^{2}}{3!} \pdd{\bF(\bx,t)}{t}\right|_{t = t^{n}} + \mathcal{O}(\dt^{3}).
    \end{split}
\end{equation}
Dropping the error term $\mathcal{O}(\dt^{3})$ in Eq.~\eqref{eq:taylor3}, we achieve
a third-order temporally accurate approximation to the averaged-flux $\bF^{avg}(\bx)$,
which is what we aimed for.

The temporal derivatives can be cast in terms of spatial derivatives
through the Cauchy–Kowalewski procedure, satisfying the governing system of equations.
We adopt an approach conceptually slightly different from~\cite{christlieb2015picard}
to convert the temporal derivatives in Eq.~\eqref{eq:taylor3}.
To wit, in~\cite{christlieb2015picard} the temporal derivatives of the fluxes
in Eq.~\eqref{eq:taylor3}
inadvertently treats $\bF$ as a function (through $\bU$) that depends only on $t$ but not on $\bx$.
The authors then followed a consistent logical pathway to convert the temporal
derivatives to the corresponding spatial derivatives via the Cauchy-Kowalewski
procedure. Although the final expression of such converted terms in~\cite{christlieb2015picard}
are found to be equivalent to our results presented below,
we introduce a new self-consistent 
approach to converting the temporal \textit{partial} derivatives  in Eq.~\eqref{eq:taylor3} to
the related spatial derivatives.

To this end, we first apply the chain rule to the governing
equation in Eq.~\eqref{eq:gov} as,
\begin{equation}\label{eq:chain-rule}
    \pd{\bU}{\bF}\pd{\bF}{t} + \pd{\bF}{x} + \pd{\bG}{y} = 0,
\end{equation}
which leads us to write an evolution equation of the $x$-flux $\bF$,
\begin{equation}\label{eq:Ft}
    \pd{\bF}{t} = -\pd{\bF}{\bU} \left( \pd{\bF}{x} + \pd{\bG}{y} \right), \quad
    \mbox{where} \quad \pd{\bF}{\bU} = \left( \pd{\bU}{\bF}\right)^{-1}.
\end{equation}
From Eq.~\eqref{eq:Ft}, the higher temporal derivatives are acquired
in a successive manner. For example, the second derivative can be computed as,
\begin{equation}\label{eq:Ftt}
    \pdd{\bF}{t} = - \pd{}{t} \left( \pd{\bF}{\bU} \right) \cdot \left( \pd{\bF}{x} + \pd{\bG}{y} \right)
    - \pd{\bF}{\bU} \cdot \pd{}{t} \left( \pd{\bF}{x} + \pd{\bG}{y} \right),
\end{equation}
where
\begin{equation}\label{eq:Jt}
    \begin{split}
        \pd{}{t} \left( \pd{\bF}{\bU} \right) = \pdd{\bF}{\bU} \cdot \pd{\bU}{t}
         = -\pdd{\bF}{\bU} \cdot \left( \pd{\bF}{x} + \pd{\bU}{y} \right),
    \end{split}
\end{equation}
and
\begin{equation}\label{eq:divt}
    \begin{split}
        \pd{}{t} \left( \pd{\bF}{x} + \pd{\bG}{y} \right) &= \pd{}{x} \left( \pd{\bF}{t} \right) +
        \pd{}{y} \left( \pd{\bG}{t} \right) \\
        & = \pd{}{x} \left( -\pd{\bF}{\bU} \cdot \left( \pd{\bF}{x} + \pd{\bG}{y} \right) \right) +
        \pd{}{y} \left( -\pd{\bG}{\bU} \cdot \left( \pd{\bF}{x} + \pd{\bG}{y} \right) \right) \\
        & = -\pdd{\bF}{\bU} \cdot \pd{\bU}{x} \cdot \left( \pd{\bF}{x} + \pd{\bG}{y} \right) -
        \pd{\bF}{\bU} \cdot \left( \pdd{\bF}{x} + \frac{\partial^{2} \bG}{\partial x \partial y} \right) \\
        &\quad -\pdd{\bG}{\bU} \cdot \pd{\bU}{y} \cdot \left( \pd{\bF}{x} + \pd{\bG}{y} \right) -
        \pd{\bG}{\bU} \cdot \left( \frac{\partial^{2} \bF}{\partial x \partial y} + \pdd{\bG}{y} \right).
    \end{split}
\end{equation}
Here, \( \pd{\bF}{\bU} \) represents an $N\times N$ flux Jacobian matrix with $\bU \in \mathbb{R}^N$,
while \( \pdd{\bF}{\bU} \) represents an $N\times N \times N$ flux Hessian tensor.
A dot product between the Hessian tensor and a vector is to be understood
as a tensor contraction~\footnote[$\ddagger$]
    {For $\bfx,\bfy \in \mathbb{R}^N$, the tensor contraction
    $\pdd{\bF}{\bU} \cdot \bfx  \cdot  \bfy $ is defined in such a way that its $k$-th component is given as
    $\Bigl[ \pdd{\bF}{\bU} \cdot \bfx  \cdot  \bfy \Bigr]_k  =
    \sum\limits_{i,j=1}^{N} \frac{\partial^2 \bF_k}{\partial \bU_i \partial \bU_j} \bfx_i \bfy_j$.},
thus a double dot products
between the Hessian tensor
and two vectors, i.e.,
\( \pdd{\bF}{\bU} \cdot \left( \; \right) \cdot \left( \; \right) \),
yields a vector of dimension $N$.

We approximate the spatial derivatives with the following 5-point central differencing formulae that are fourth-order,
\begin{equation}\label{eq:dx}
    \pd{\bF_{i}}{x} = \frac{\bF_{i-2} - 8 \bF_{i-1} + 8 \bF_{i+1} - \bF_{i+2} }{12\dx} + \mathcal{O}(\dx^{4}),
\end{equation}
\begin{equation}\label{eq:dxx}
    \pdd{\bF_{i}}{x} = \frac{-\bF_{i-2} + 16 \bF_{i-1} - 30 \bF_{i} + 16 \bF_{i+1} - \bF_{i+2} }{12\dx^{2}} + \mathcal{O}(\dx^{4}).
\end{equation}
Theoretically speaking, in Eq.~\eqref{eq:taylor3},
we are able to retain the third-order temporal accuracy for $\bF^{avg}(\bx)$
as long as we approximate
${\partial \bF}/{\partial t} = \mathcal{O}(\dx^2)$ and
${\partial^2 \bF}/{\partial t^2} = \mathcal{O}(\dx)$,
since in a hyperbolic system the Courant condition guarantees $\dt = \mathcal{O}(\dx)$.
In this regard, our design choice with the fourth-order approximations in
Eqs.~\eqref{eq:dx} and~\eqref{eq:dxx} is indeed surfeited.
In addition, between Eqs.~\eqref{eq:dx} and~\eqref{eq:dxx},
it is sufficient to retain one order lower in Eq.~\eqref{eq:dxx} than in Eq.~\eqref{eq:dx}
because of the extra $\dt$ factor for the second derivative term in Eq.~\eqref{eq:taylor3}.
In fact, in case with ${\partial^2 \bF_i}/{\partial x^2} = \mathcal{O}(\dx^2)$,
some of our test results have indicated that the convergence rate drops slightly
without affecting the expected third-order solution accuracy.
In practice, the use of any lower order approximation in Eq.~\eqref{eq:dx} or Eq.~\eqref{eq:dxx}
does not gain the overall computational performance; hence we stick to the fourth-order
approximations.

For the cross derivatives, 
the simple second order central differencing is sufficient to retain the third-order temporal accuracy,
\begin{equation}\label{eq:dxy}
    \frac{\partial^{2} \bF}{\partial x \partial y} =
    \frac{\bF_{i+1, j+1} - \bF_{i-1, j+1} - \bF_{i+1, j-1} + \bF_{i-1 j-1}}{4\dx\dy} + \mathcal{O}(\dx^{2}, \dy^{2}).
\end{equation}
The use of the finite central differencing formulae above leads us to
approximate all the terms described in Eq.~\eqref{eq:taylor3} with sufficient accuracy,
and the time-averaged flux \( \bF^{avg} \) is ready to be reconstructed by the WENO5 method.
The \( y \)-directional flux \( \bG^{avg} \) can be obtained in a similar way, and
this finalizes the general PIF procedure introduced in~\cite{christlieb2015picard}.

What remains to discuss is the system-free methodology, an approach
newly introduced in Section~\ref{sec:system-free},
to provide a computationally affordable and efficient strategy for Jacobian and Hessian
calculations. As can be seen in~\cite{christlieb2015picard},
the straight analytical calculations of the Jacobian matrix
\( \pd{\bF}{\bU} \) and the Hessian tensor \( \pdd{\bF}{\bU} \)  make
hurdles to implement the PIF into a code in practice.
The situation becomes even worse when extending
the current method to a fourth-order or higher,
in which case higher derivatives will require high dimensional tensors
whose size, $N^D$ with $D>3$, grows dramatically.
%
Although the Jacobian and the Hessian calculations
can be easily obtained with the aid of symbolic
manipulators such as \texttt{SymPy}, \texttt{Mathematica}, or \texttt{Maple},
it still demands complicated coding/debugging efforts and ample memory consumption.
Furthermore, as the Jacobian/Hessian calculations highly depend
on the type of the governing system under consideration,
it is required to re-derive the Jacobian/Hessian terms analytically
every time we need to solve a new system,
e.g., shallow water equations, or magnetohydrodynamics (MHD) equations,
to name a few.

With this in mind, in the following section, we propose a new way to attain a high-order
approximation for the time-averaged fluxes without the need for direct
calculation of Jacobians or Hessians.
The method is to be considered as a system-free PIF method, which will be referred to as
SF-PIF in what follows.

\section{System-Free Approach}\label{sec:system-free}

Our goal in this section is to provide a new alternate formulation of computing
the multiplications of Jacobian-vector and Hessian-vector-vector terms in Eqs.~\eqref{eq:Ftt}~--~\eqref{eq:divt}.
The new approach will replace the necessity for analytical derivations of
these Jacobian and Hessian terms  in the original PIF method that
are system-dependent,
with a new system-independent formulation, based on
the so-called ``Jacobian-free'' method which
is widely used for Newton-Krylov-type iterative
schemes~\cite{brown1990hybrid, knoll2004jacobian, knoll2011application, gear1983iterative}.
We also extend it to a ``Hessian-free'' method.
The resulting PIF method, called the system-free PIF (SF-PIF),
approximates the Jacobian and Hessian terms via
the Jacobian-free and Hessian-free approach.

\subsection{Jacobian approximation}\label{subsec:jac-approx}
To begin with our discussion,
we consider a Taylor expansion for the flux vector at a small displacement from \( \bU \),
\begin{subequations}\label{eq:FeV}
    \begin{align}
        \label{eq:FeV-right}\bF ( \bU + \varepsilon \bV) =
        \bF(\bU) + \varepsilon \pd{\bF}{\bU} \cdot \bV +
        \frac{1}{2} \varepsilon^{2} \pdd{\bF}{\bU} \cdot \bV \cdot \bV + \mathcal{O}(\varepsilon^{3}), \\
        \label{eq:FeV-left}\bF ( \bU - \varepsilon \bV) =
        \bF(\bU) - \varepsilon \pd{\bF}{\bU} \cdot \bV +
        \frac{1}{2} \varepsilon^{2} \pdd{\bF}{\bU} \cdot \bV \cdot \bV + \mathcal{O}(\varepsilon^{3}),
    \end{align}
\end{subequations}
where \( \bV \) is an arbitrary vector that has
the same number of components as \( \bU \), and \( \varepsilon \) is a
small scalar perturbation.
By subtracting Eq.~\eqref{eq:FeV-left} from Eq.~\eqref{eq:FeV-right},
we get an expression of a central differencing that is of second-order in $\varepsilon$,
\begin{equation}\label{eq:jac-free}
    \pd{\bF}{\bU} \cdot \bV =
    \frac{\bF(\bU + \varepsilon \bV) - \bF(\bU - \varepsilon \bV)}{2\varepsilon}
    + \mathcal{O}(\varepsilon^{2}).
\end{equation}

Alternatively, the first-order forward differencing or the backward differencing can be used here.
However, we choose the above second-order central differencing
so that the order of accuracy of the entire system-free approach consistently
scales with \( \mathcal{O}(\varepsilon^{2}) \),
given that
the Hessian approximation described in the next subsection
is to be bounded by \( \mathcal{O}(\varepsilon^{2}) \).
With the system-free approximation of Jacobian, 
all the Jacobian-vector products 
in
Eqs.~\eqref{eq:Ft}~--~\eqref{eq:divt} are to be replaced
with the central differencing in Eq.~\eqref{eq:jac-free}.

\subsection{Hessian approximation}\label{subsec:hes-approx}
The flux Hessian tensor is unavoidable in designing the third-order PIF method in time.
Below, we derive two different approximations with
$\mathcal{O}(\varepsilon^{2})$ accuracy
to estimate the two different types of the tensor appearing in
Eqs.~\eqref{eq:Ftt}--\eqref{eq:divt}.
Namely, in the first type the Hessian tensor contracts with the same vector, e.g., $\bV$, twice,
and in the second type the tensor contracts with two different vectors, e.g., $\bV$ and $\bW$.

For the first type, we use a Taylor expansion analogous to
Eq.~\eqref{eq:FeV} to approximate the
Hessian-vector-vector product with a central differencing of order $\mathcal{O}(\varepsilon^{2})$,
\begin{equation}\label{eq:hes-free}
    \pdd{\bF}{\bU} \cdot \bV \cdot \bV =
    \frac{\bF(\bU + \varepsilon \bV) - 2 \bF(\bU) + \bF(\bU - \varepsilon \bV)}{\varepsilon^{2}}
    + \mathcal{O}(\varepsilon^{2}).
\end{equation}
Using a simple vector calculus,
the second type can be derived from the first type in Eq.~\eqref{eq:hes-free}
by exploring a symmetric property of the Hessians,
\begin{equation}\label{eq:hes-v-w}
    \pdd{\bF}{\bU} \cdot \bV \cdot \bW =
    \frac{1}{2} \left( \pdd{\bF}{\bU} \cdot \left( \bV + \bW \right) \cdot \left( \bV + \bW \right) -
    \left( \pdd{\bF}{\bU} \cdot \bV \cdot \bV + \pdd{\bF}{\bU} \cdot \bW \cdot \bW \right) \right)
    + \mathcal{O}(\varepsilon^{2}).
\end{equation}
The Hessian approximations derived here are now ready to be substituted
in Eqs.~\eqref{eq:Ftt}--\eqref{eq:divt}.

\subsection{Proper choices of $\varepsilon$}\label{subsec:epsilon}

In the previous subsections, we approximated
the Jacobian-vector product and the Hessian-vector-vector product
with a small perturbation \( \varepsilon \).
The choice of \( \varepsilon \) has to be considered carefully
as it affects the solution accuracy and stability.
On one hand, we need to minimize \( \varepsilon \)
to improve the approximated solution accuracy,
the quality of which will scale as the truncation error of
$\mathcal{O}(\varepsilon^{2})$.
On the other hand, if it is too small
the solution would be contaminated by
the floating-point roundoff error which is bounded by
the machine accuracy
\( \varepsilon_{\text{mach}} \)~\cite{knoll2004jacobian}.
Therefore, $\varepsilon$ is to be determined judiciously
to provide a good balance between the two types of error.

A recent study by \textit{An et al.}~\cite{an2011finite}
presents an effective analysis of choosing
\( \varepsilon \) in the context of the Jacobian-free Newton-Krylov iterative framework.
The authors have shown how to compute an ideal value of
\( \varepsilon \) which minimizes the error of the
central differencing in the Jacobian-vector approximation.
We follow the same idea to obtain ideal values of \( \varepsilon \)
for each Jacobian-vector and Hessian-vector-vector approximation.

The main idea in~\cite{an2011finite} is to find a good
balance between the truncation error \( \mathcal{O}(\varepsilon^{2}) \)
of each Jacobian-free approximation in Eq.~\eqref{eq:jac-free}
and Hessian-free approximation in Eq.~\eqref{eq:hes-v-w},
and the intrinsic floating-point roundoff error $\delta\bF(\bx)$ when calculating
the target exact function value $\bF(\bx)$ with
an approximate value $\bF(\bx) + \delta\bF(\bx)$.
The perturbation $\delta\bF(\bx)$ may include any errors characterized
in computer arithmetic such as roundoff errors, and is assumed to be
bounded by the machine accuracy,
i.e., $|| \delta\bF(\bx) || \le \varepsilon_{\text{mach}}\sim 2.2204 \times 10^{-16}$.
In this section, we only display the outcome of this analysis which will provide a good optimal estimation
of $\varepsilon$. More details of the relevant study are found in~\cite{an2011finite,knoll2004jacobian}.

Let \( \varepsilon^{op} \) denote the optimal value of \( \varepsilon \), and in particular, we consider
computer arithmetic in a 64-bit machine with double precision.
Following the analysis in~\cite{an2011finite}, we obtain \( \varepsilon^{op} \) for each
Jacobian-free and Hessian-free approximation,
\begin{equation}\label{eq:sigma}
    \begin{split}
        \varepsilon^{op}_{\text{jac}} &= \argmin_{\sigma > 0}
        \left( \frac{\sigma^{2}}{2} + \frac{\varepsilon_{\text{mach}}}{2 \sigma} \right) =
        {\left( \frac{\varepsilon_{\text{mach}}}{2} \right)}^{\frac{1}{3}} \approx \num{4.8062e-06},\\
        \varepsilon^{op}_{\text{hes}} &= \argmin_{\sigma > 0}
        \left( \frac{\sigma^{2}}{3} + \frac{\varepsilon_{\text{mach}}}{\sigma^{2}} \right) =
        {\left( 3 \varepsilon_{\text{mach}} \right)}^{\frac{1}{4}} \approx \num{1.6065e-4},
    \end{split}
\end{equation}
where \( \varepsilon^{op}_{\text{jac}} \) and \( \varepsilon^{op}_{\text{hes}} \)
represent the optimal
$\varepsilon$
value for the Jacobian-free and Hessian-free approximations, respectively.

However, direct use of \( \varepsilon^{op} \) as
the displacement step size in the central differencing schemes
in Eq.~\eqref{eq:jac-free} and Eq.~\eqref{eq:hes-v-w}
is not a good idea for stability reasons.
Usually, the vector \( \bV \) in
Eq.~\eqref{eq:jac-free} and Eq.~\eqref{eq:hes-free} could have an
enormous value in a strong shock region, so it is safer to use
a smaller step size to preserve the needed stability. To meet this,
we normalize the ideal value, \( \varepsilon^{op} \),
by the magnitude of the vector \( \bV \).
There are several prescriptions available
in the Jacobian-free Newton–Krylov
literatures~\cite{knoll2004jacobian, brown1990hybrid}
to help finalize our decision of choosing a proper value of \( \varepsilon \)
as a function of  \( \varepsilon^{op} \).
Nonetheless, we have seen that,
in a suite of test problems in Section~\ref{sec:results},
a simple approach of
taking a square root of \( \varepsilon^{op} \)
with a simple normalization is sufficient to attain the desired accuracy and stability,
which is given as,
\begin{equation}\label{eq:eps}
    \overline{\varepsilon} = \frac{\sqrt{\varepsilon^{op}}}{\left\lVert \bV \right\rVert_{2}}.
\end{equation}

Lastly, we complete the \( \varepsilon \) estimation by
taking the minimum value between $\overline{\varepsilon}$ and $\dt$,
\begin{equation}\label{eq:min_eps_dt}
    \begin{split}
        \varepsilon = \min \left( \overline{\varepsilon}, \; \dt  \right),
    \end{split}
\end{equation}
considering the fact that
the Jacobian-free and Hessian-free terms
will be multiplied by (at least) \( \Delta t \) after all.
As a consequence, the choice of $\varepsilon$ in Eq.~\eqref{eq:min_eps_dt}
ensures the third-order temporal accuracy in the overall SF-PIF scheme.

\section{Results}\label{sec:results}
In this section, we present numerical results of the SF-PIF algorithm
on a suite of well-known benchmark problems in 1D and 2D
using two different systems of equations. We first demonstrate
the test results of SF-PIF on the system of 1D and 2D Euler equations,
and compare them with
the test results of RK3 and the original PIF method.
The SF-PIF is also tested on the system of 2D shallow water equations
to illuminate the system-free nature of the SF-PIF algorithm.

\subsection{1D Euler equations}\label{subsec:1d-euler}
In this section we test SF-PIF to simulate numerical problems
in 1D Euler equations, defined as,
\begin{equation}\label{eq:1d-gov}
    \pd{\bU}{t} + \pd{\bF(\bU)}{x} = 0,
\end{equation}
where the conservative variables and the corresponding fluxes are given as,
\begin{equation}\label{eq:1d-euler}
    \bU =
    \begin{bmatrix}
        \rho \\
        \rho u \\
        E
    \end{bmatrix}, \quad
    \bF (\bU) =
    \begin{bmatrix}
        \rho u \\
        \rho u^{2} + p \\
        u \left( E + p \right)
    \end{bmatrix}.
\end{equation}

We adopt the spatially fifth-order WENO5 method~\cite{jiang1996efficient} for all test problems,
combined with three
different time integrators: the traditional three-stage
third-order Runge-Kutta method (RK3)~\cite{gottlieb1998total},
the original third-order Picard Integration Formulation (PIF)~\cite{christlieb2015picard}, and
our third-order system-free PIF (SF-PIF) method presented in this paper.
In all three combinations the nominal solution accuracy is expected to be
fifth-order in space \( \mathcal{O} (\dx^{5}) \)
and third-order in time \( \mathcal{O} (\dt^{3}) \).
We use a fixed Courant number, \( \text{C}_{\text{cfl}} = 0.7 \) in all 1D cases.

\subsubsection{Sine wave advection}\label{sec:1Dsine_advection}

We start with a simple convergence test to see if the desired solution accuracy
is retrieved in smooth flows.
To that end, we choose a simple passive advection
on a smooth flow configuration
following the same setup as in~\cite{lee2017piecewise}.
The density profile is initialized with a sinusoidal wave,
\( \rho (x) = 1.5 - 0.5\sin(2\pi x)\).
The $x$-velocity and
the pressure are set as constant values of \( u = 1\) and
\( p = 1/\gamma \)
with the specific heat ratio, \( \gamma = 5/3 \).
Albeit solved using the nonlinear Euler equations, the problem is
solved in a linear regime, viz., the velocity and pressure remain
constant for all $t\ge 0$ so that the initial sinusoidal density profile
is purely advected by the constant velocity $u=1$ without any nonlinear
dynamics such as a formation of shocks and rarefactions.

\begin{figure}[ht!]
    \centering
    \includegraphics[width=100mm]{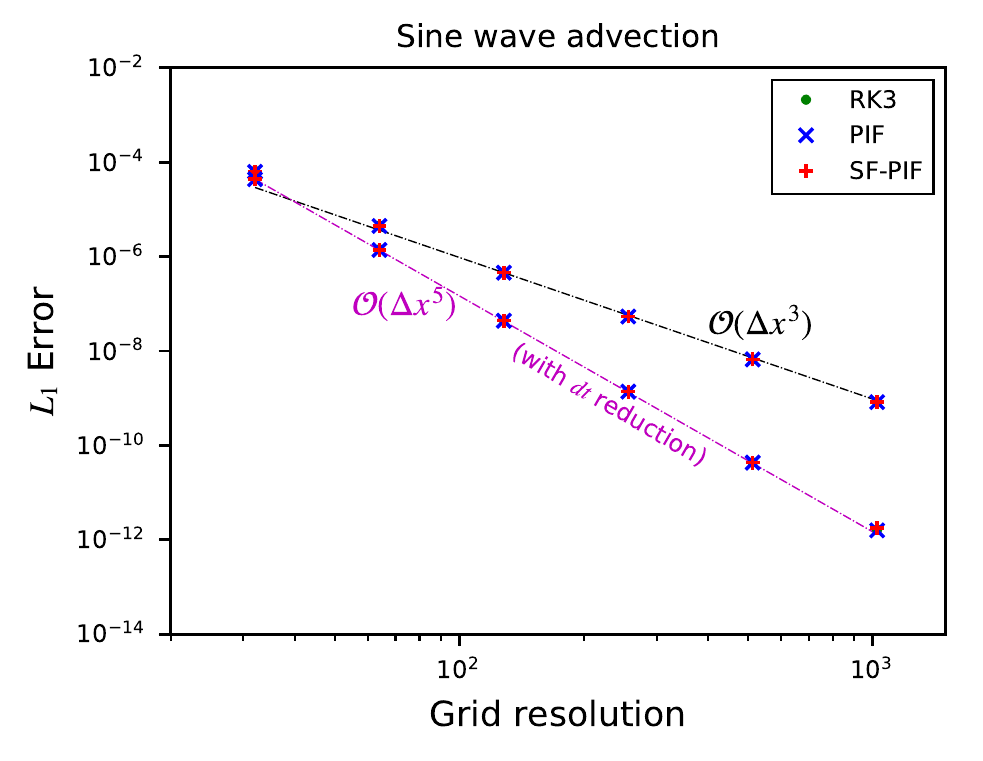}
    \caption{Convergence test for the 1D sine wave advection problem.
        The errors are calculated
        in \( L_{1} \) sense against the initial density profile
        resolved on the computational grids refined
        from 32 to 1024 by a factor of 2.
        All numerical solutions follow the theoretical third-order convergence rate
        (the black-dotted line) when using the timesteps
        computed from the Courant condition.
        Also plotted are the solutions of using reduced timesteps, which follows the
        fifth-order convergence rate represented in the pink-dotted line.
    }\label{fig:1d-order}
\end{figure}

The simulation domain is defined on an interval \( [0, 1] \)
with the periodic boundary condition on the both ends.
The density profile will propagate for one period through
the simulation domain and will return to its initial position
at \( t = 1 \).
On return, by being a linear problem,
any deformation of the density profile from the initial shape
can be considered as a numerical error
associated with phase errors or numerical diffusions.
The accuracy of the numerical solutions is measured
by computing the \( L_{1} \)
error between the solution at \( t = 1 \)
and the initial profile
%
on a different number of cells
\( N_{x} = 32, 64, 128, 256, 512\) and \( 1024 \).
The results are
depicted in Fig.~\ref{fig:1d-order} for
all three different temporal integrating schemes, RK3, PIF, and SF-PIF\@.

There are two types of result demonstrated in Fig.~\ref{fig:1d-order}.
In the first type, the discrete solutions of three different temporal schemes
advanced with timesteps computed from the Courant condition with $C_{\text{cfl}}=0.7$.
Interestingly, we observe that the numerical solutions from SF-PIF
as well as the other two schemes converge at third-order,
indicating that
the overall
leading error of the simulation is dominated by the third-order accuracy of the
temporal schemes despite the use of the more accurate fifth-order
spatial WENO5 solver.
As also briefly mentioned in Section~\ref{sec:introduction}, at low-mid resolutions,
solution accuracy of simulations is mostly governed by the spatial accuracy
until the leading error of the solution is caught up by the temporal error as
computational grids get further refined to higher resolutions.
So, in this problem,
one may expect that the numerical solutions would pick up a faster convergence,
faster than third-order and closer to fifth-order, at least at the lower end of
the grid resolutions.
Unfortunately, this doesn't happen in this test and the third-order
temporal accuracy quickly takes over the control throughout the entire range of
the grid resolutions tested herein. This solution behavior supports strongly the importance
of integrating spatially interpolated/reconstructed solutions with
a temporal scheme whose accuracy is sufficiently high enough to be
well comparable to that of the spatial solver.

In the second type, however, timesteps are restricted in order to
match up the lower third-order temporal accuracy of $\mathcal{O}(\Delta t^3)$
with the higher fifth-order spatial accuracy of $\mathcal{O}(\Delta x^5)$~\footnote[$\S$]
    {In numerical PDEs, it is a common practice to combine a spatial solver
    whose accuracy is higher than the accuracy of a temporal solver, with
    an exception of ADER schemes in which
    the same order of accuracy is preserved.
    }.
We follow the usual trick of timestep reduction (e.g., see~\cite{mignone2010high}) to
manually adjust the timestep $\Delta t_N$ on a grid size of $N$  to
an adjusted value, satisfying the equal rate of change between the
spatial and temporal variations.
As a result, the restricted (or reduced) timestep is defined by, 
\begin{equation}\label{Eq:dt_reduction_a}
    {\Delta t_N} = {\Delta t_0} \Big( \frac{\Delta x_N}{\Delta x_0} \Big)^{\frac{5}{3}}.
\end{equation}
The sub-indices ``$0$'' and ``$N$'' refer to the
time and grid scales on a nominal coarse and a fine resolution, respectively.
For instance, in the current configuration,
$\dx_0$ is the grid scale on $N_x=32$ and
$\dt_0$ is the corresponding timestep subject to the Courant condition with $C_{\text{cfl}}=0.7$.
%
With the reduction, the overall leading error of the simulation
(from both spatial and temporal)
is matched with the fifth-order spatial accuracy of WENO5, and
the numerical solution of SF-PIF follows the fifth-order rate as expected.
Moreover, the SF-PIF solution compares indistinctly well with the other two solutions,
almost in the same pattern.
Even though the timestep reduction helps improve the numerical convergence rate,
particularly when a temporal solver has a lower accuracy than a spatial solver,
such simulations suffer not only from a much longer computational time to reach a target
final time $t=t_{\max}$ but also from a degradation of solution effectiveness
in resolving small scales due to an extended
simulation time~\cite{kent2014determining_part1,kent2014determining_part2}.

In all cases, the solution of SF-PIF compares equally well with the solutions of the original PIF
and RK3 in both quantitatively and qualitatively.

\subsubsection{Sod's shock tube problem}

The Sod's shock tube problem~\cite{sod1978survey} is the one of the most famous hydrodynamics
test problems for testing a numerical scheme's capability to handle discontinuities and shocks.
The initial condition is given as,
\begin{equation}\label{eq:sod-init}
    \left( \rho, u, p \right) = \begin{cases}
        \left( 1, 0, 1 \right) & \text{for } x \le 0.5, \\
        \left( 0.125, 0, 0.1 \right) & \text{for } x > 0.5,
    \end{cases}
\end{equation}
in a simulation box of \( [0, 1] \), with outflow boundary conditions
on both ends at \( x = 0 \) and \( x = 1 \).

\begin{figure}[ht!]
    \centering
    \begin{subfigure}{80mm}
        \centering
        \includegraphics[width=\textwidth]{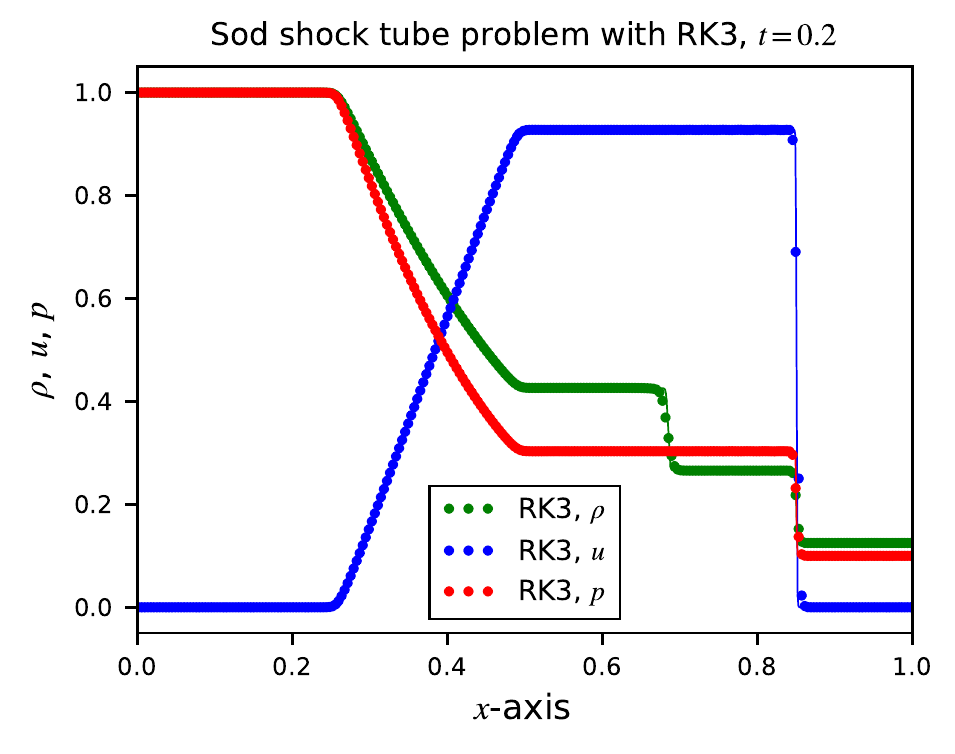}
        \caption{}\label{subfig:sod_rk3}
    \end{subfigure}
    \begin{subfigure}{80mm}
        \centering
        \includegraphics[width=\textwidth]{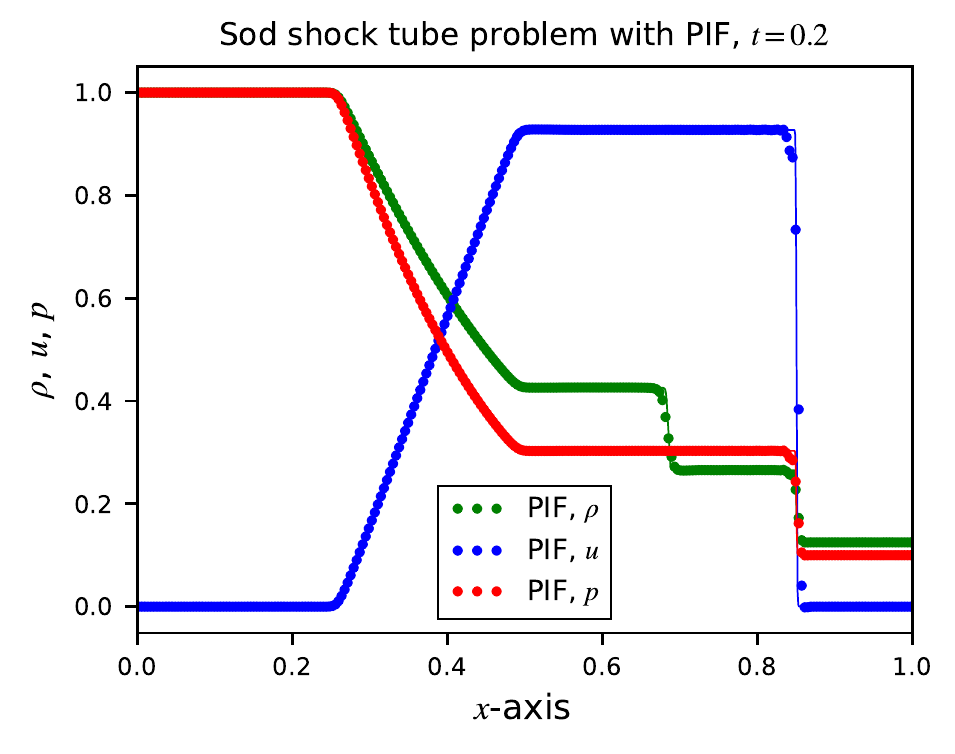}
        \caption{}\label{subfig:sod_pif}
    \end{subfigure}
    \begin{subfigure}{80mm}
        \centering
        \includegraphics[width=\textwidth]{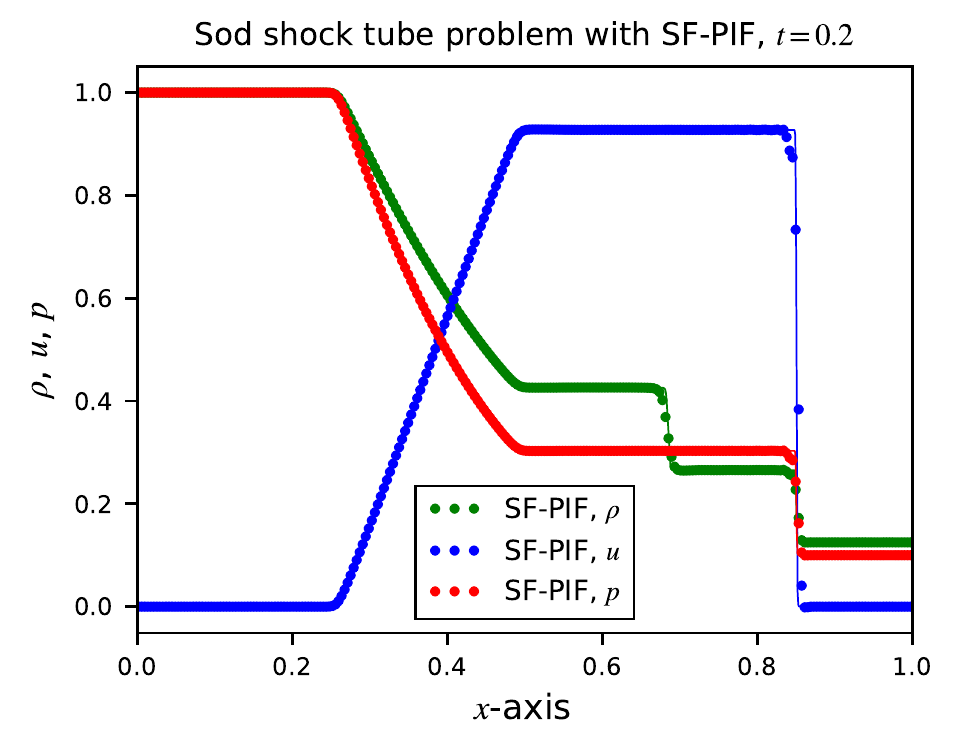}
        \caption{}\label{subfig:sod_sf3}
    \end{subfigure}
    \caption{Sod's shock tube problem at \( t = 0.2 \). The reference solutions are over-plotted
        as solid lines in each panel, which are resolved on a grid resolution of \( N_{x} = 1024 \)
        with RK3. The symbols in each panel represent the solution resolved on
        \( N_{x} = 256 \) grid cells with
        (\protect\subref{subfig:sod_rk3}) RK3,
        (\protect\subref{subfig:sod_pif}) PIF, and (\protect\subref{subfig:sod_sf3}) SF-PIF\@.
    }\label{fig:1d-sod}
\end{figure}

The results with the grid size of \( N_{x} = 256 \) at \( t = 0.2 \) are plotted as symbols in Fig.~\ref{fig:1d-sod}.
The solid lines on each panel represent the reference solution resolved on a more finer grid size,
\( N_{x} = 1024 \), by using WENO5+RK3.
%
As seen in Fig.~\ref{fig:1d-sod}, the solutions of SF-PIF agree with the reference solution
and the two solutions from RK3 and PIF\@. We see that,
in PIF and SF-PIF,
there is a slight oscillation in the $x$-velocity immediately behind the shock front.
We believe that this small oscillation is
originated from the use of the central differencing formulae in
Eqs.~\eqref{eq:dx}~--~\eqref{eq:dxy} for both SF-PIF and PIF\@.
Even though we haven't found any test case in our numerical experiments that
this oscillatory prediction negatively impacts the
overall performance of the SF-PIF scheme,
this could potentially degrade the capability of shock/discontinuity
handling in SF-PIF in more extreme problems. As such, this issue will be
investigated further in our future studies.

\subsubsection{Two-blast wave}
Since originally introduced by Woodward and Colella~\cite{woodward1984numerical},
this test problem has been chosen
by many practitioners to
examine codes' capability of capturing the correct dynamics of
shock-shock interactions.
Initially, two strong shocks are developed at each end of the simulation box \( \left[ 0, 1 \right] \).
The reflecting boundary conditions are used at both ends, $x=0$ and $x=1$.
They are driven to each other,
leading to a highly compressive collision
at the middle part of the domain at $t\sim 0.028$ by the initial pressure discontinuities.
The shock-shock collision produces an extremely high and narrow density peak.
One of the main points of interest is to demonstrate how well numerical solutions
predict the density profile and its peak amplitude
particularly at \( t = 0.038 \),
one hundredth of a second after the strong collision.
\begin{figure}[ht!]
    \centering
    \begin{subfigure}{80mm}
        \centering
        \includegraphics[width=\textwidth]{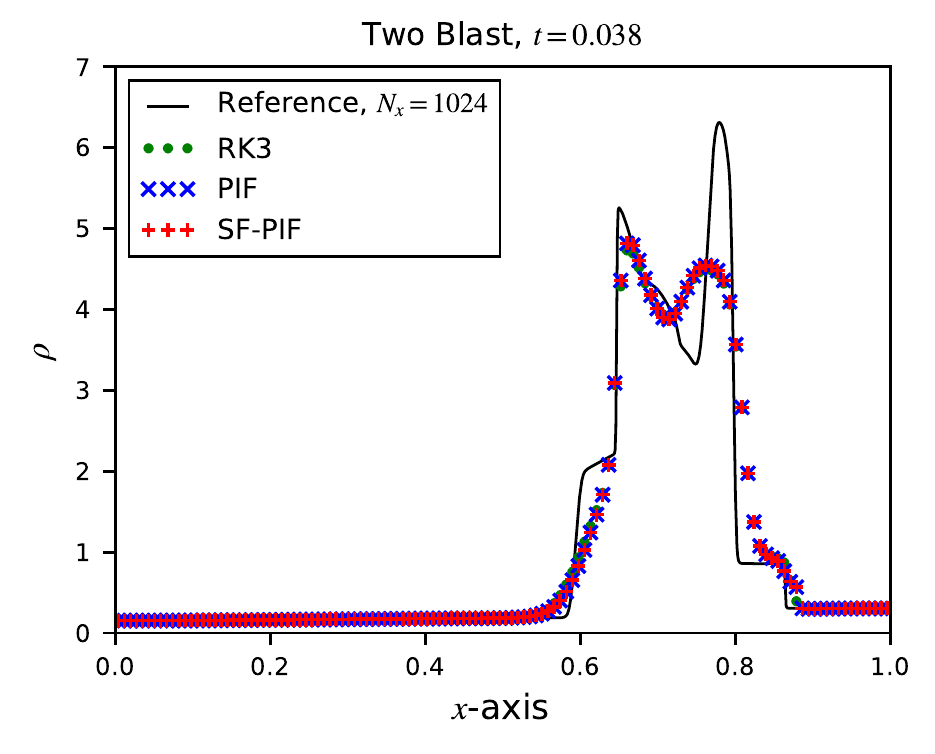}
    \end{subfigure}
    \begin{subfigure}{80mm}
        \centering
        \includegraphics[width=\textwidth]{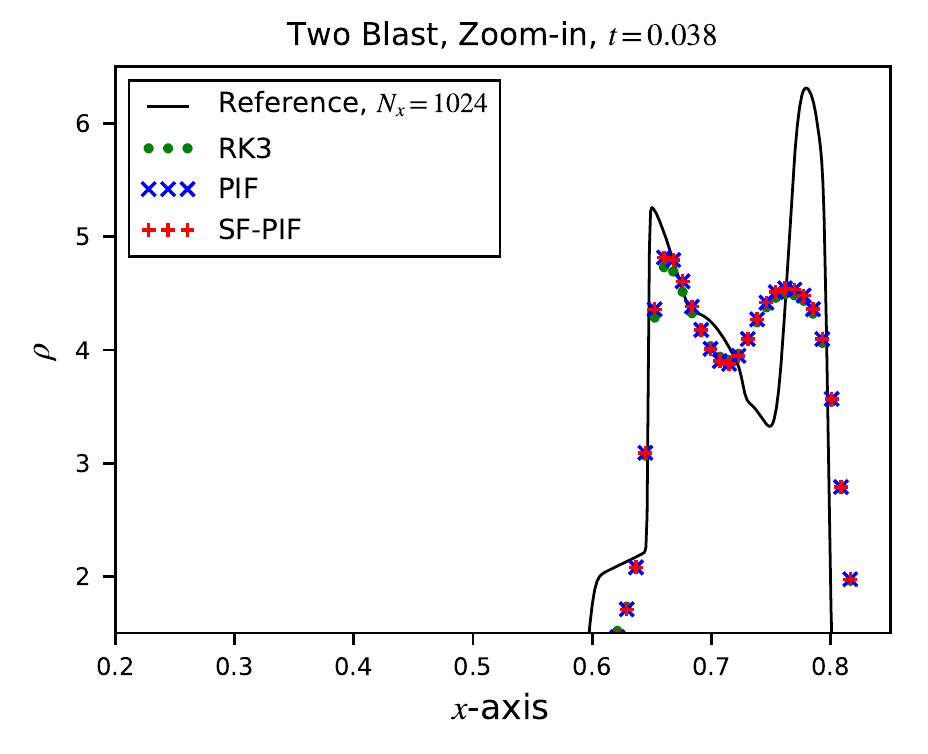}
    \end{subfigure}
    \caption{Two-blast wave problem at \( t = 0.038 \).
        The symbols depict the numerical solutions resolved on
         a resolution of \( N_{x} = 128 \),
        and the solid line represents the reference solution on a finer grid, \( N_{x} = 1024 \),
        with RK3 method. Right: A close-up view of the shock interaction region.
    }\label{fig:1d-blast}
\end{figure}

In Fig.~\ref{fig:1d-blast}, the density profile with SF-PIF is plotted against two other density profiles
with RK3 and PIF on a grid resolution of  \( N_{x} = 128 \).
Over-plotted is the reference solution (the black solid curve)
integrated with RK3 on a grid size \( N_{x} = 1024 \) for comparison.
We see that all three methods produce an acceptable quality of solutions
and they are all in good agreement.
As observed in the close-up view in the right panel of Fig.~\ref{fig:1d-blast},
there are slight improvements in the solutions of SF-PIF and PIF over
the RK3 solution
in resolving the peak density amplitudes following the sharp gradients
over $0.63 \le x \le 0.8$.
It is not surprising to see that
the two solutions are nigh equivalent
since the two methods share the same common
ground in mathematical formulation.
On the other hand, the equivalency of the two methods affirms that
our system-free approach described in Section~\ref{sec:system-free}
not only provides ease of code implementation but also guarantees
high fidelity in solution accuracy compared against
the analytical approach of the original PIF method.

\subsubsection{Shu-Osher problem}

The Shu-Osher problem~\cite{shu1989efficient} describes an interaction of a Mach 3 shock
and a smooth density profile. Initially, a right-going Mach 3 shock propagates into the low
density profile with sinusoidal perturbations. As the shock proceeds, it is superposed with the
sinusoidal density profile, leaving two different trails of density behind the right-propagating shock.
\begin{figure}[ht!]
    \centering
    \begin{subfigure}{80mm}
        \centering
        \includegraphics[width=\textwidth]{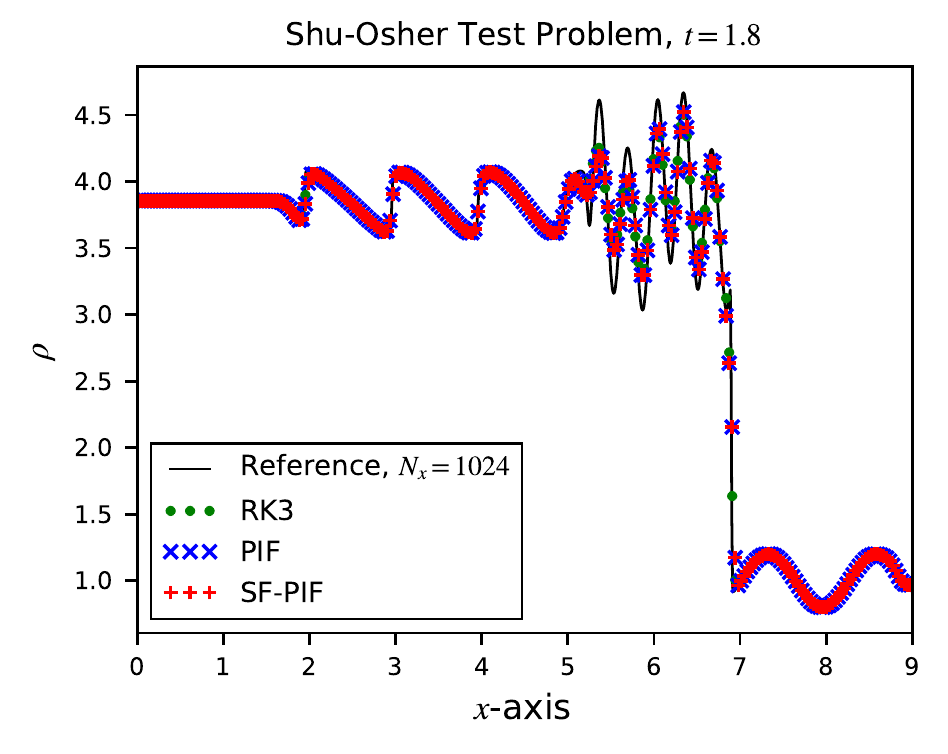}
    \end{subfigure}
    \begin{subfigure}{80mm}
        \centering
        \includegraphics[width=\textwidth]{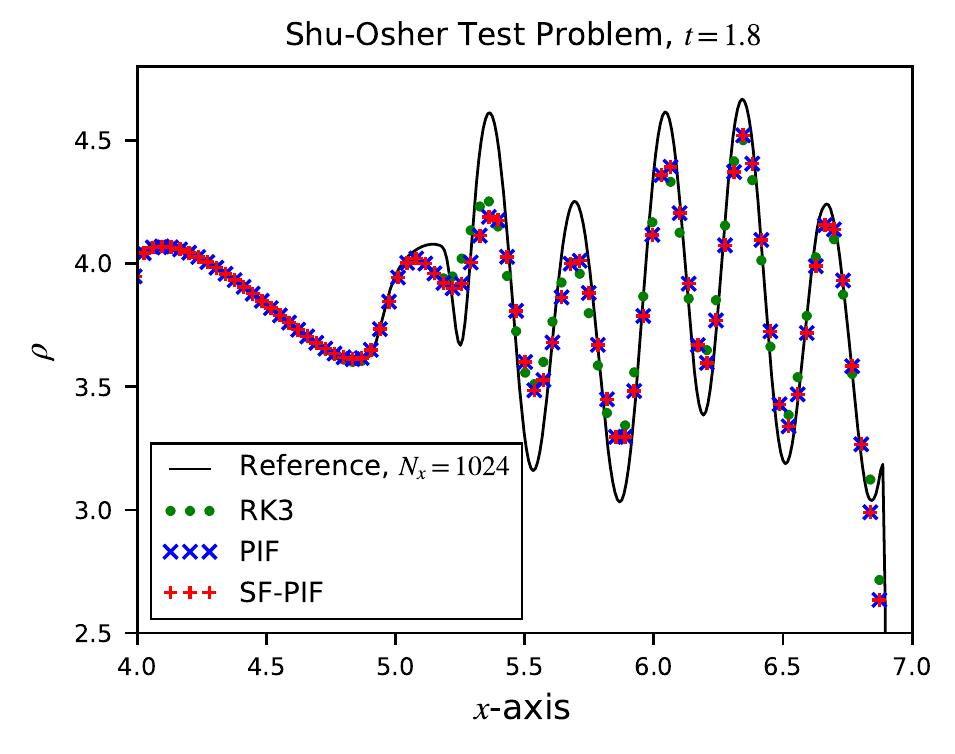}
    \end{subfigure}
    \caption{The Shu-Osher test problem at \( t = 1.8 \).
        The reference solution, plotted as a black solid curve,
        is obtained on \( N_{x} = 1024 \) grid resolution
        with RK3 method.
        Other solutions are resolved on \( N_{x} = 256 \) grid resolution,
        represented as symbols.
        Right: A close-up view of the doubled frequency region to illustrate methods' numerical diffusivity.
    }\label{fig:1d-shu}
\end{figure}
The initial sine wave gets compressed by the penetrating shock, doubling the perturbation frequency
in the downstream of the immediate post-shock region.
From the tail of the high frequency region to the farther left,
the perturbation returns to the original frequency, but at the same time,
the profile experiences
a shock-steepening which leads the initial smooth profile
into a sequence of sharp density gradients further away to the downstream.

Our main focus on this test problem is to check whether the SF-PIF scheme
is able to advance the discrete solutions in a stable and accurate manner,
capturing both the smooth and discontinuous profiles
in the double-frequency region as well as at the shock front.
Fig.~\ref{fig:1d-shu} shows the SF-PIF result at \( t = 1.8 \) with the other two methods of
RK3 and PIF\@. All three results are resolved on \( N_{x} = 256 \) grid cells,
while the reference solution is obtained on a much higher resolution of \( N_{x} = 1024 \)
with RK3 method, displayed in a black solid curve.
The results illustrated in the close-up view on the right show that the two single-stage methods,
SF-PIF and PIF, are able to produce density peak amplitudes in the
double-frequency region slightly higher than the solution from the
multi-stage RK3 method. As before, the two solutions of SF-PIF and PIF are
found to be almost indistinguishable, assuring the validity of the system-free
formulations of the Jacobian and Hessian terms in SF-PIF
compared with the analytic calculations of the same terms in the original PIF scheme.
%
%

\subsection{2D Euler equations}\label{subsec:2d-euler}

In this section, we implement the SF-PIF scheme to solve the two-dimensional Euler equations,
defined by Eq.~\eqref{eq:gov} with,
\begin{equation}\label{eq:2d-euler}
    \bU = \begin{bmatrix}
        \rho \\
        \rho u \\
        \rho v \\
        E
    \end{bmatrix},\quad
    \bF (\bU) = \begin{bmatrix}
        \rho u \\
        \rho u^{2} + p \\
        \rho u v \\
        u \left( E + p \right)
    \end{bmatrix}, \quad
    \bG (\bU) = \begin{bmatrix}
        \rho v \\
        \rho u v \\
        \rho v^{2} + p \\
        v \left( E + p \right)
    \end{bmatrix}.
\end{equation}

As before, WENO5 is used as the spatial method
in all 2D test problems in this section
so that we can focus on differing numerical effects only from the temporal methods.
We present results from several 2D test problems
and compare them to address the performance of the SF-PIF scheme.
We choose the CFL number \( C_{\text{cfl}} = 0.4 \) in
all problems.

\subsubsection{Nonlinear isentropic vortex advection}\label{sec:vortex}
Our first 2D problem considers SF-PIF's accuracy and speedup
in two-dimensional Euler equations.
We conduct the nonlinear isentropic vortex advection problem
originally introduced by Shu~\cite{shu1998essentially}.
Here, we modify the problem as presented in~\cite{spiegel2015survey}
where the size of the periodic domain $[0, 20] \times [0, 20]$ is twice larger in each direction
than the original setup in~\cite{shu1998essentially}
to prevent self-interactions of the vortex across the periodic domain.
The same consideration was also found in~\cite{lee2017piecewise,reyes2019variable}.
Initially positioned at the center of the domain,
the isentropic vortex starts to advect in the positive diagonal direction
and returns to the initial location after one periodic time, $t=20$.
The \( L_{1} \) error is calculated through the direct
comparison between the initial condition
and the numerical solution at $t=20$ to examine numerical convergence rates
on four different grid resolutions, $N_x=N_y=50, 100, 200$, and $400$.

\begin{table}[hb!]
    \centering
    \caption{The \( L_{1} \) errors, the rates of convergence,
    and the relative computation times for the vortex advection test.
    Here, we display the comparison between RK3 and SF-PIF only
    since the difference between SF-PIF and PIF is indistinguishable.
    }\label{table:vortex}
    \begin{tabular}{@{}llccllcc@{}}
%
        \toprule
        \multirow{2}{*}{\( N_{x} = N_{y} \)} & RK3 &  &  &  & \multicolumn{3}{l}{SF-PIF} \\
        \cmidrule(lr){2-4} \cmidrule(l){6-8}
        &   \(L_{1}\) error & \(L_{1}\) order   & Speedup &  & \(L_{1}\) error & \(L_{1}\) order & Speedup \\ \midrule
         50  & \num{7.22E-1} & \---              & 1.0 &  & \num{6.95E-1}   & \---              & 0.64 \\
        100 & \num{5.76E-2} & 3.65            & 1.0 &  & \num{5.58E-2}   & 3.64            & 0.56 \\
        200 & \num{2.94E-3} & 4.29            & 1.0 &  & \num{2.89E-3}   & 4.27            & 0.56 \\
        400 & \num{1.22E-4} & 4.59            & 1.0 &  & \num{1.26E-4}   & 4.52            & 0.51
    \end{tabular}
\end{table}

\begin{figure}[ht!]
    \centering
    \begin{subfigure}{80mm}
        \centering
        \includegraphics[width=0.95\textwidth]{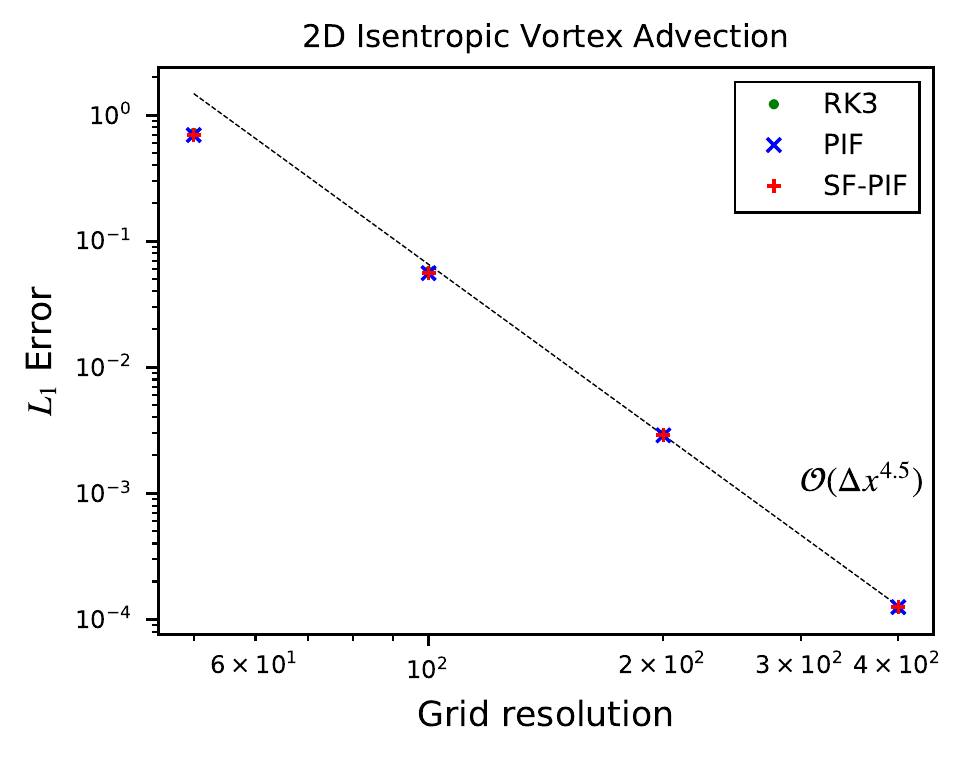}
    \end{subfigure}
    \begin{subfigure}{80mm}
        \centering
        \includegraphics[width=0.95\textwidth]{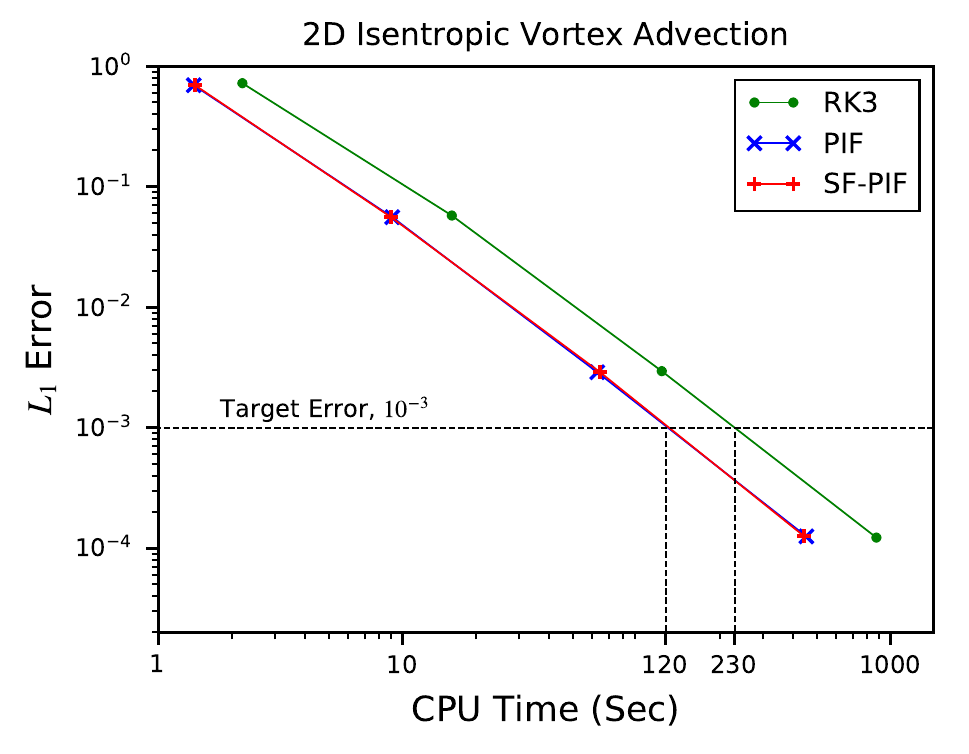}
    \end{subfigure}
    \caption{Left: The \( L_{1} \) errors of the isentropic vortex advection problem
        with respect to the grid resolutions, \( 50, 100, 200, \) and \( 400 \).
        All simulations use WENO5 for spatial reconstruction with CFL=0.4.
        Right: The \( L_{1} \) errors of the isentropic vortex advection problem
       as a function of computational time.
    }\label{fig:vortex}
\end{figure}

The results of the convergence test are summarized in Table~\ref{table:vortex}
and depicted on the left panel in Fig.~\ref{fig:vortex}.
We see a good comparable match in magnitudes of the $L_1$ errors
across all three methods.
Between SF-PIF and PIF, there is
no significant distinction in their accuracy and performance, as demonstrated
on both panels
in Fig.~\ref{fig:vortex}. All three methods converge at
4.5th order towards the initial condition at each grid resolution.
Unlike the 1D linear sine advection test in Section~\ref{sec:1Dsine_advection},
we do not see the full dominance in error from the third-order temporal discretization
(at least over the range of the grid resolutions tested herein),
which could potentially reduce the overall convergence rate down
to third-order as in the sine advection case.
Concurrently, we also see that the solution does not converge at
full fifth-order either, the rate of which is due from the use of WENO5.
This can be explained as a nonlinear effect in which
the overall leading error term of the fifth-order spatial discretization
is slightly compromised by the lower third-order time integration schemes.
In general, an exact analysis of this phenomenon is highly problem dependent
and it often becomes intractable in many nonlinear cases.
In practice though, this issue can be improved by one of the two approaches,
timestep reductions, or developments of fourth-order or higher temporal methods.
As discussed, the first option is not favorable in most large-scale simulations, and hence
the latter is to be considered as a better alternative from
the mathematical standing point.

On the right panel, we illustrate the performance of SF-PIF
measured in terms of the CPU time to reach a target error threshold, say
$L_1$ error of $10^{-3}$.
As demonstrated, the results of the two single-stage methods,
SF-PIF and PIF, coincide to each other and their performance is
approximately twice faster than the multi-stage method of RK3.
It is worth to be noted that
the SF-PIF approach can be readily swappable with
an RK integrator in an existing code
without too much effort,
leaving any existing spatial implementations intact mostly.
Moreover, such a code transformation with SF-PIF is more advantageous
in simplicity than with the original PIF method
because SF-PIF replaces PIF's need for analytic
derivations of the Jacobian and Hessian terms with
the system-free approximations, which have shown to be
highly commensurate with the analytical counterparts of the PIF scheme.
%

\subsubsection{2D Riemann problems}\label{sec:2drp}
Next, we consider 2D Riemann problems which are extensively
studied in~\cite{zhang1990conjecture, schulz1993classification, schulz1993numerical}
and widely used for the code verifications~\cite{lee2017piecewise, balsara2010multidimensional, buchmuller2014improved, don2016hybrid,reyes2019variable}.
Among a wide range of different configurations,
we demonstrate 
two test cases,
Configuration 3 and Configuration 5,
following the descriptions
in~\cite{don2016hybrid,lee2017piecewise}.
Both test cases are resolved on a square simulation box of
\( \left[ 0, 1 \right] \times \left[ 0, 1 \right] \)
with \( 400 \times 400 \) grid cells using outflow boundary conditions.

\begin{figure}[ht!]
    \centering
    \begin{subfigure}{80mm}
        \centering
        \includegraphics[width=0.95\textwidth]{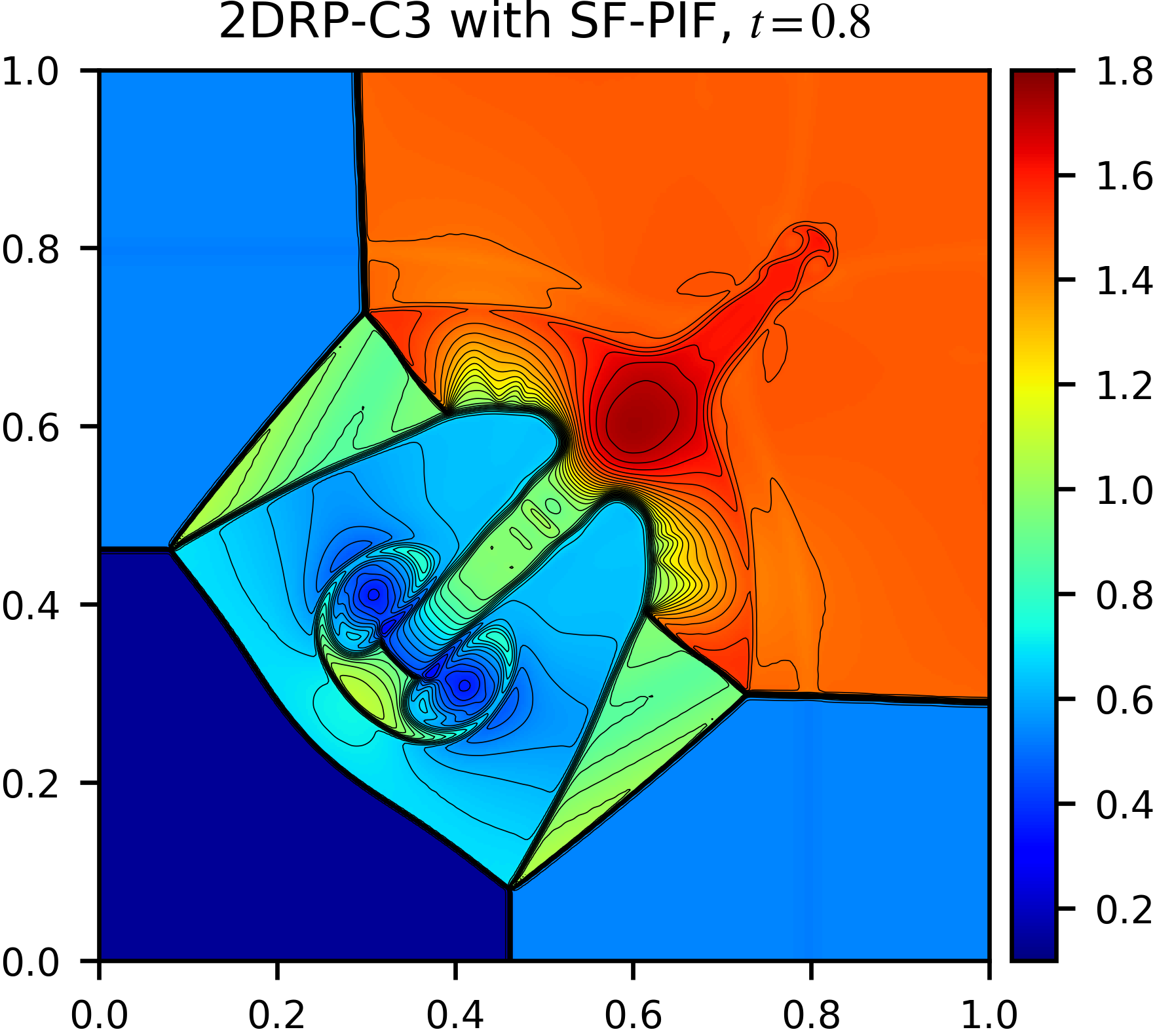}
    \end{subfigure}
    \begin{subfigure}{80mm}
        \centering
        \includegraphics[width=0.95\textwidth]{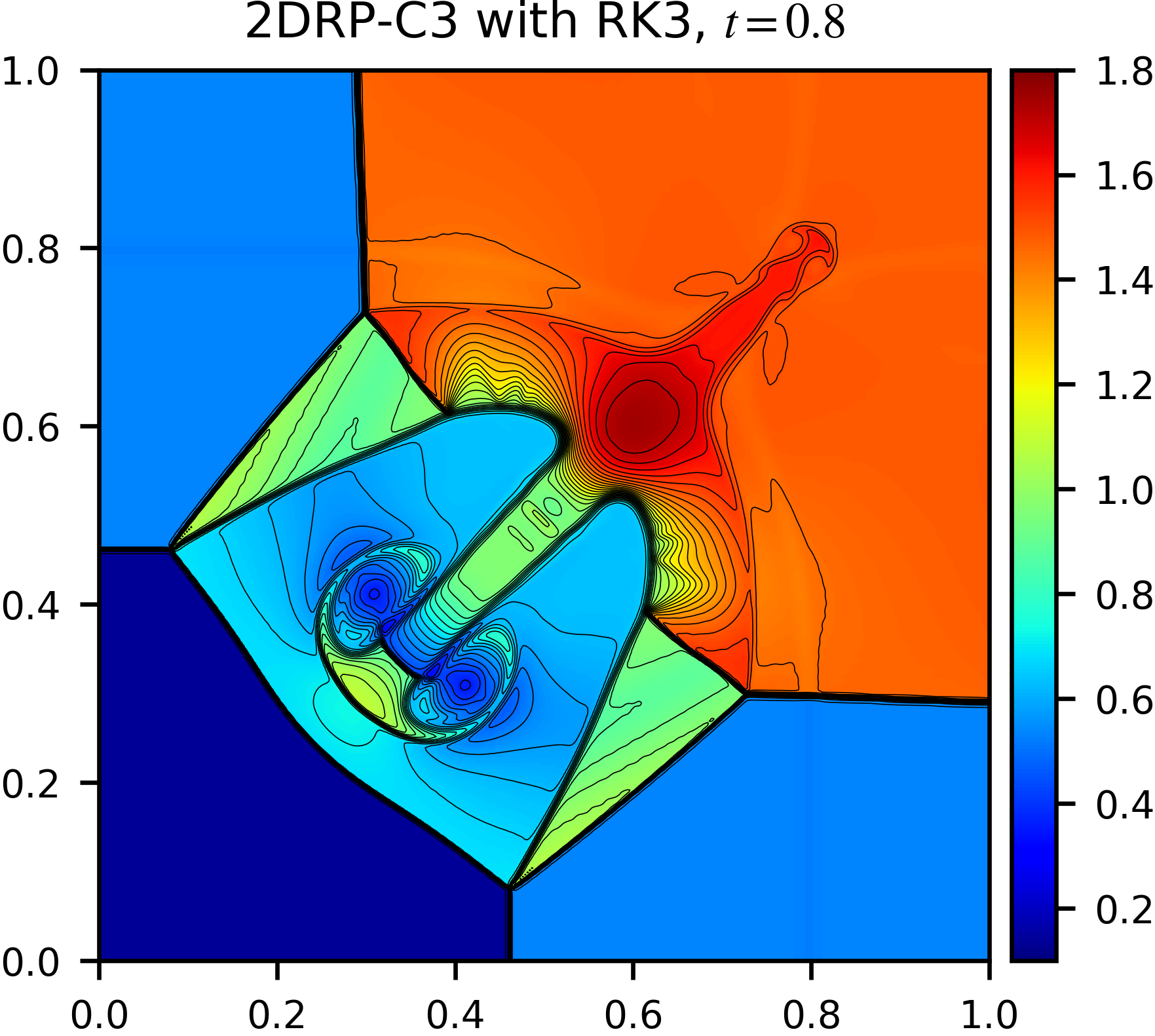}
    \end{subfigure}
    \caption{The density profile of Configuration 3
        with SF-PIF (left) and with RK3 (right).
        The color map ranges from \( 0.1 \) to \( 1.8 \), and
        40 evenly-spaced contour lines are over-plotted with
        the same range.
    }\label{fig:2drp_c3}
\end{figure}

\begin{figure}[ht!]
    \centering
    \begin{subfigure}{80mm}
        \centering
        \includegraphics[width=0.95\textwidth]{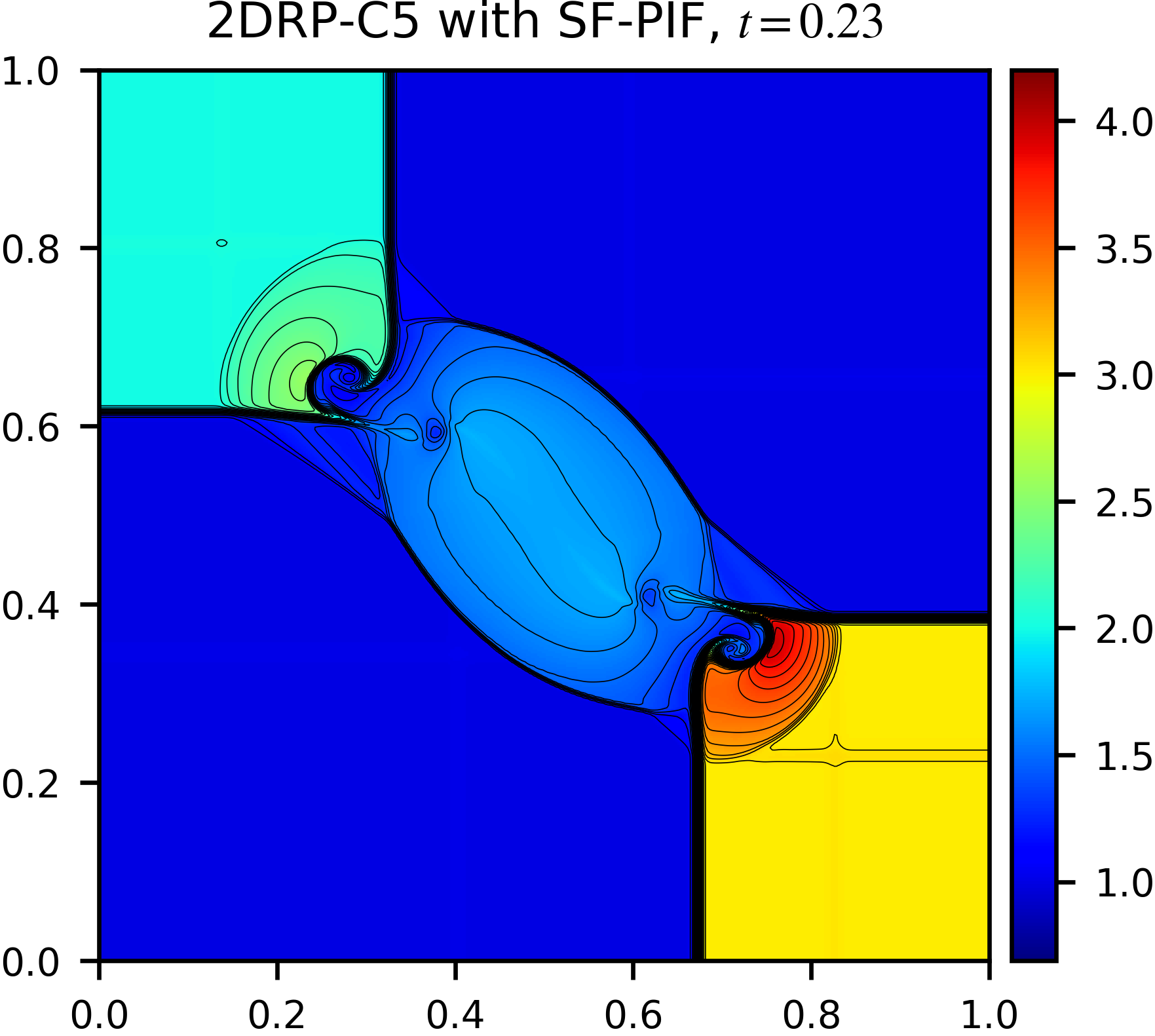}
    \end{subfigure}
    \begin{subfigure}{80mm}
        \centering
        \includegraphics[width=0.95\textwidth]{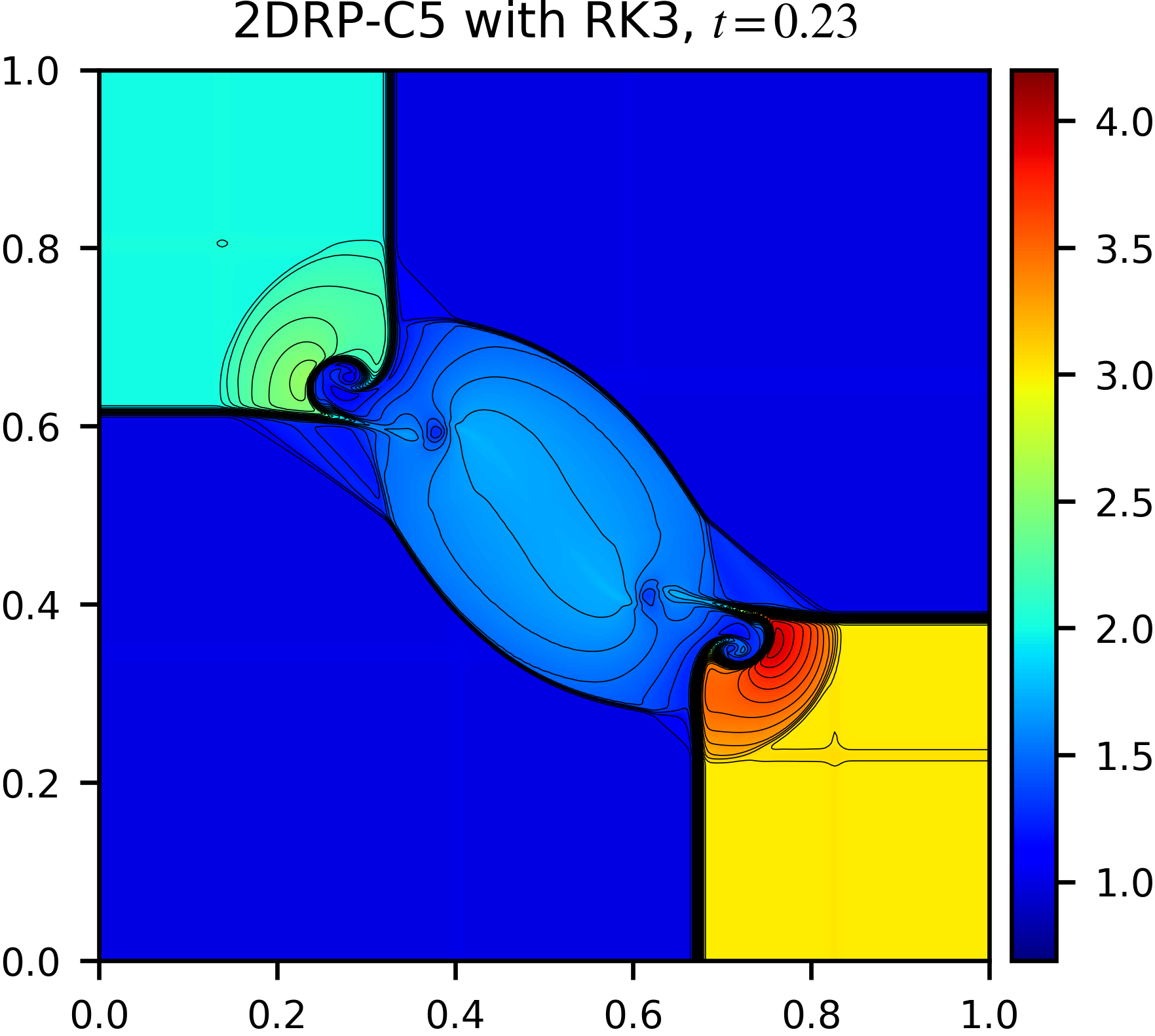}
    \end{subfigure}
    \caption{The density profile of Configuration 5
        with SF-PIF (left) and with RK3 (right).
        The color map ranges from \( 0.69 \) to \( 4.2 \), and
        40 evenly-spaced contour lines are over-plotted with
        the same range.
    }\label{fig:2drp_c5}
\end{figure}

The results of Configuration 3 and Configuration 5 are shown in Fig.~\ref{fig:2drp_c3}
and Fig.~\ref{fig:2drp_c5}, respectively.
The density maps are visualized with
pseudo-color schemes ranging between 0.1 and 1.8 in Configuration 3,
and between 0.69 and 4.2 in Configuration 5.
Over-plotted are
40 levels of contour lines using the same density range
in each configuration.

We see that the SF-PIF result is well comparable to that of RK3,
which assures that SP-PIF method is eminently reliable
in temporally advancing solutions with
discontinuities and shocks in 2D,
while at the same time, preserving the expected symmetries
and flow profiles in each problem.
In both, however, there is yet no formation of secondary
Kelvin-Helmoltz instabilities, typically exhibited as vortical rollups
along the slip lines~--~the regions connecting
the inner edges of the triangle ``arms''
surrounding the mushroom-jet
and the stem of the mushroom-jet.
Direct comparisons with other FDM test results on
the same grid resolution with comparable accuracy
(e.g., Fig. 12 in~\cite{reyes2019variable}; Fig. 6 in~\cite{don2016hybrid})
reveal that their test results have more pronounced
rollup structures developed along the slip lines.
This implies that both of the test results here have
more numerical diffusivity than those approaches therein~\cite{don2016hybrid,reyes2019variable}.
We first emphasize that developments of small-scale features
relevant to grid scales depend much more sensitively on the choice of
spatial solvers (e.g., WENO5 with the global Rusanov flux-splitting in our case),
but less on a temporal solver.
It is well-known that the global Rusanov flux-splitting
(used in this study but not in~\cite{don2016hybrid,reyes2019variable})
has added numerical diffusivity, which can delay an onset of the secondary
instability formation in our experiments.
Secondly, with mathematical rigor,
an increasing number of rollups in any given scheme should only be
understood as a proof of a \textit{less} amount of numerical dissipation,
but not as a proof of a \textit{better} numerical method, than others.
Measuring a \textit{proper} amount of such rollups in any given simulations
on a set of specific setups (i.e., grid resolutions, spatial and temporal
orders of numerical methods, parameters, etc.)
is not a simple task. Such a study will need to involve an extensive
systemic comparison analysis that requires
careful validation and verification tests,
the topics of which are beyond the scope of this paper.

\subsubsection{Implosion test}\label{sec:implosion}

The next problem to consider is the implosion test problem
introduced by Hui \textit{et al.}~\cite{hui1999unified}.
An unsteady flow configuration is given as an initial condition which launches
a converging shock wave towards the domain center.
We follow a simpler version by Liska and Wendroff~\cite{liska2003comparison},
which takes only the right upper quadrant
of the original setup in~\cite{hui1999unified}
as the simulation domain.
In this setup, the simulation is initialized on a region of a square domain,
\( \left[ 0, 0.3 \right] \times \left[ 0, 0.3 \right] \),
enclosed with reflecting walls,
in which case a converging shock wave is launched toward the lower left corner
at $(x,y)=(0,0)$.
The initial shock wave gets bounced by the reflecting walls and produces a double Mach
reflection along two edges of $x=0$ and $y=0$. As a consequence,
two jets are formed along the edges moving toward the origin $(x,y)=(0,0)$ and collide
each other. This two-jet collision then ejects a newly-formed jet into the diagonal direction
$x=y$. Reflecting shocks continuously interact with the diagonal jet, turning it into
a longer and narrower shape over time. The observed structures of filaments and fingers
along the diagonal jet and at its base are progressively intensified by the
Ritchmyer-Meshkov instability, a level of which depends sensitively on
numerical dissipation.
%
%
The shape of the jet is the key view point of
the implosion test since it is a good indicator of
a code's symmetric property and numerical dissipation.
If the numerical scheme fails to maintain a high level of symmetry,
the jet will eventually be derailed off-diagonally and deformed over time.
Besides, an excess amount of numerical dissipation will
turn the jet into a less narrow and less elongated shape along the diagonal.

\begin{figure}[ht!]
    \centering
    \begin{subfigure}{80mm}
        \centering
        \includegraphics[width=0.95\textwidth]{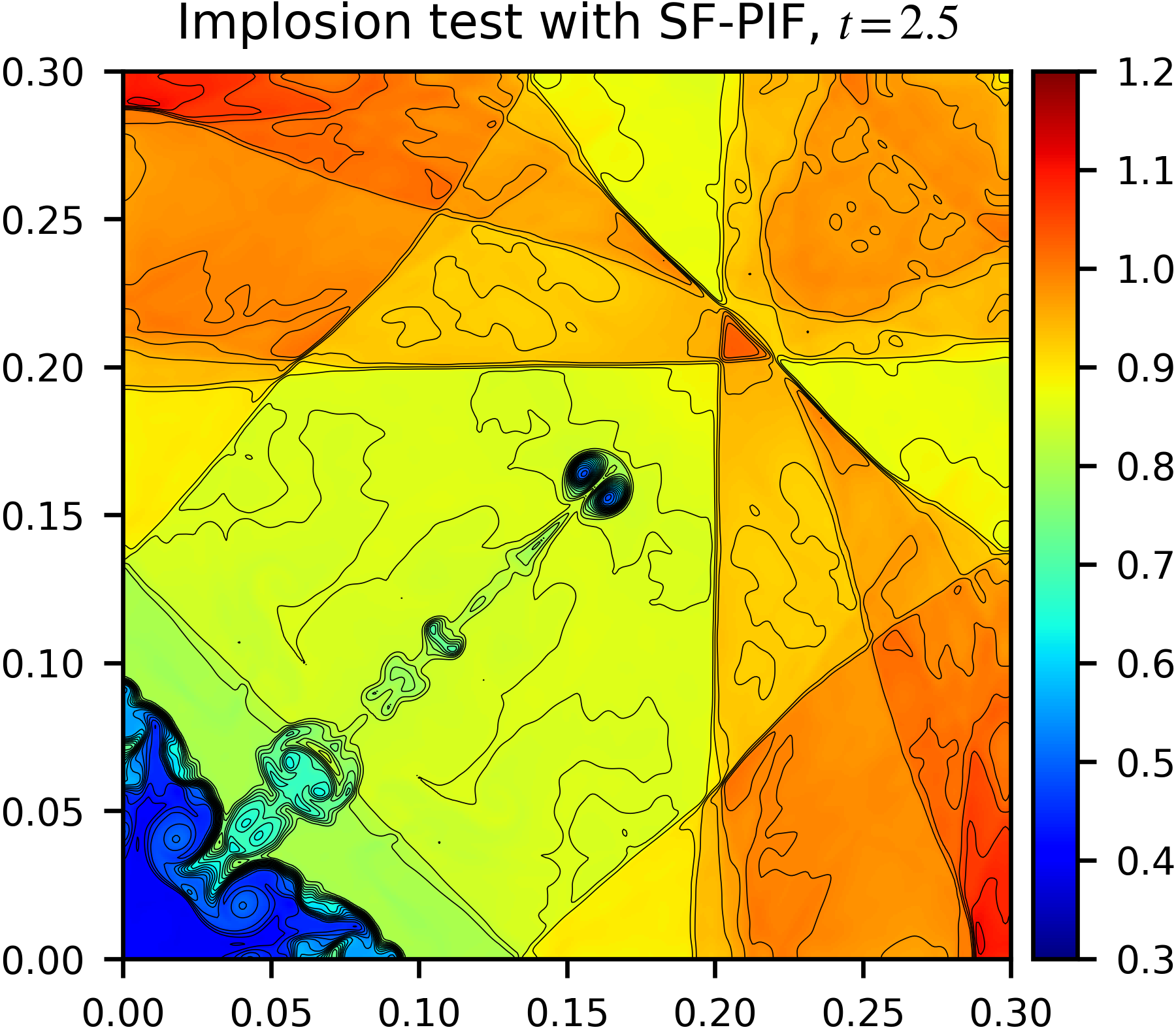}
    \end{subfigure}
    \begin{subfigure}{80mm}
        \centering
        \includegraphics[width=0.95\textwidth]{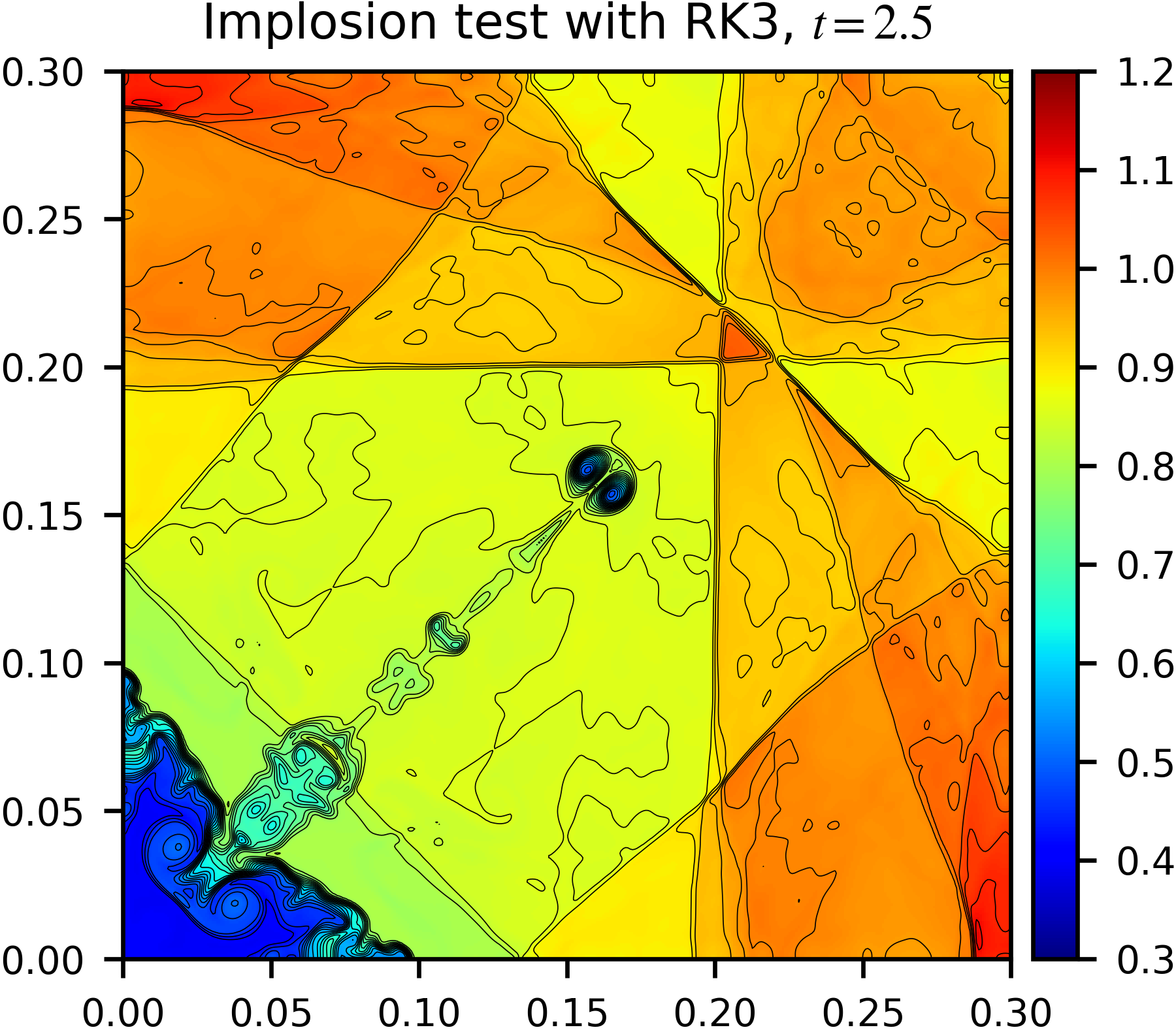}
    \end{subfigure}
    \caption{The density profile of the implosion test
        with SF-PIF (left) and with RK3 (right).
        The color map ranges from \( 0.3 \) to \( 1.2 \), and
        40 evenly-spaced contour lines are over-plotted with
        the same range.
    }\label{fig:implosion}
\end{figure}

We display our results on a \( 400 \times 400 \) grid resolution at $t=2.5$.
The result with SF-PIF is on the left panel in Fig.~\ref{fig:implosion} and
RK3 on the right.
These results can also be directly compared with
Fig.~4.7 in~\cite{liska2003comparison} and Fig.~17 in~\cite{stone2008athena}.
We clearly see that the contour lines
of SF-PIF (as well as RK3)
are shown to retain the diagonal symmetry at a highly sufficient level.
At the same time, the shape of the diagonal jet using SF-PIF matches
well with the shape using RK3,
and hence is suffice to demonstrate that the numerical dissipation in SF-PIF
is well-managed compared with RK3.

\subsubsection{Double Mach reflection}\label{sec:dmr}

\begin{figure}[hb!]
    \centering
    \begin{subfigure}{160mm}
        \centering
        \includegraphics[width=0.90\textwidth]{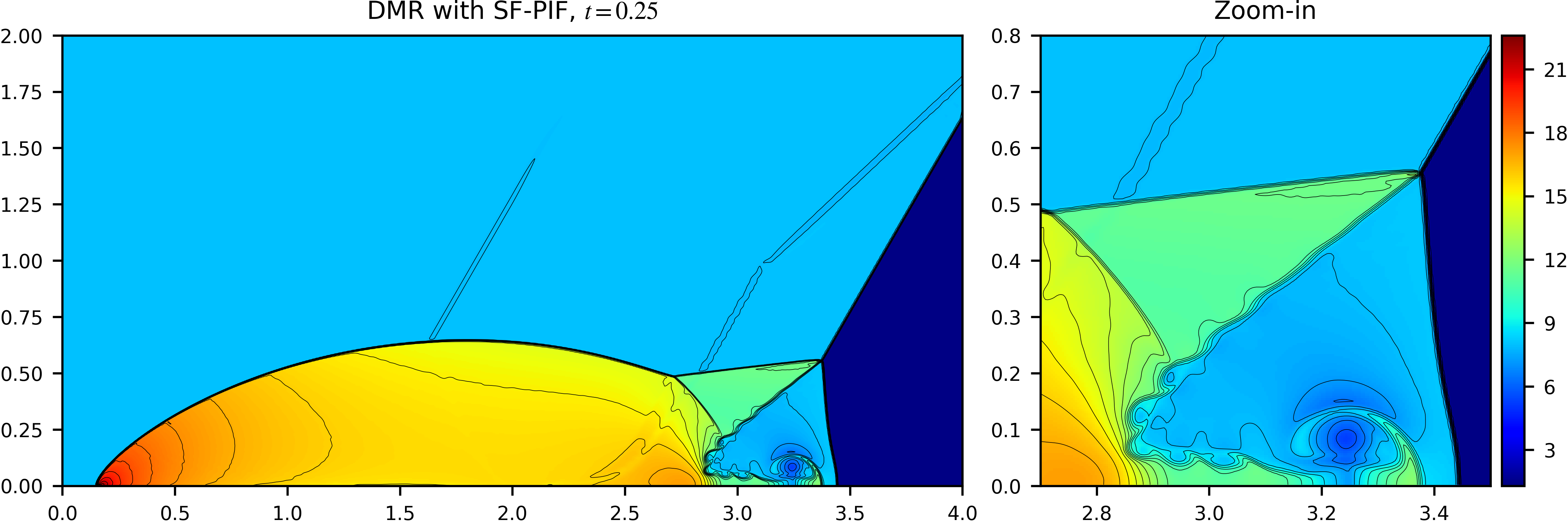}
    \end{subfigure}\vspace{3mm}
    \begin{subfigure}{160mm}
        \centering
        \includegraphics[width=0.90\textwidth]{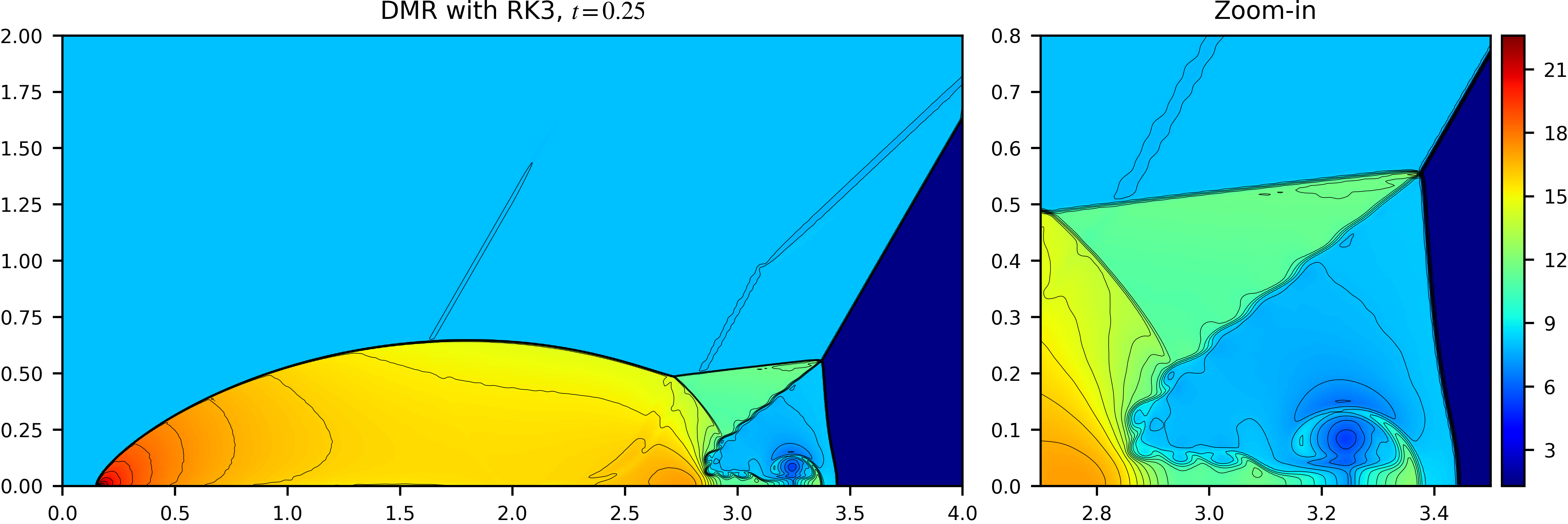}
    \end{subfigure}
    \caption{Left column: The density profile of the double Mach reflection (DMR)
        problem solved with SF-PIF (top) and RK3 (bottom).
        Right column: Close-up views near the triple-point of the density profile.
        The contour lines display 30 linearly separated points.
        All plots use the same range of density, \( [1.3, 22.6] \).
        All calculations are conducted on a grid resolution of
        \( 1024 \times 512 \) with $C_{\text{cfl}}= 0.4$.}\label{fig:dmr}
\end{figure}

Our last test using the Euler equations in 2D
is the double Mach reflection problem introduced by
Woodward and Colella~\cite{woodward1984numerical}.
We follow the original setup except that we doubled the domain size
in the \( y-\)direction to avoid any disturbances by the numerical artifacts
from the top boundary~\cite{kemm2016proper}.
A Mach 10 planar shock is initialized at the left side of the domain
with a $30^{\circ}$ angle to the reflecting bottom surface.
As the shock propagates to the right, the bottom wall continuously
bounces off the shock wave and creates a round reflected shock.
The solution further evolves into forming
two contact discontinuities and two Mach stems,
as well as a jet along the bottom surface.
The formation of this jet is similar to the formation of the two jets
in the previous implosion test,
the collision of which led to the upward moving diagonal jet.

The main point of discussion is to observe
the appearance of the contact discontinuity
that spans from the triple-point at $(x,y)\approx (3.38,5.5)$
in Fig.~\ref{fig:dmr} to the dense jet along the bottom wall,
displayed in the panels on the right column
in Fig.~\ref{fig:dmr}.
A lower amount of numerical
dissipation in a scheme makes
this region more susceptible to
Kelvin-Helmholtz instability, leading to an onset of
vortical roll-ups along the slip line.
Henceforth, the amount of such roll-ups serves
as an indicator of the scheme's numerical diffusivity.
As discussed in the 2D Riemann problems in Section~\ref{sec:2drp},
we once more point out that
such an assessment should only be used to address each scheme's
numerical diffusivity, but not to be used to rank different schemes
without careful validation and verification studies.

The density results are represented in Fig.~\ref{fig:dmr},
and the detailed views are shown
in the right panel.
Shown in the top row in Fig.~\ref{fig:dmr} is the density profile
integrated using SF-PIF
in the range of \( [1.3, 22.6] \) at $t=0.25$,
computed on a grid resolution of \( 1024 \times 512 \).
We also plot 30 evenly spaced contour lines of density in the
same range. The result with RK3 is in the bottom row.
We can see that the time prediction of density using SF-PIF matches
sufficiently well with the result using RK3,
confirming the validity of SF-PIF in comparison.
%
%

\subsection{Shallow water equations}\label{subsec:2d-shallow}
In this section, we switch the governing equations to a system of
2D shallow water equations (SWE) without a source term, defined
by a conservative hyperbolic system as,
\begin{equation}\label{eq:2d-shallow}
    \bU = \begin{bmatrix}
        h \\
        h u \\
        h v
    \end{bmatrix},\quad
    \bF (\bU) = \begin{bmatrix}
        h u \\
        h u^{2} + \frac{1}{2} g h^{2} \\
        h u v
    \end{bmatrix}, \quad
    \bG (\bU) = \begin{bmatrix}
        h v \\
        h u v \\
        h v^{2} + \frac{1}{2} g h^{2}
    \end{bmatrix}.
\end{equation}
Here, \( h \) is the vertical depth of the fluid,
\( \mathbf{v} = \left( u, v \right)  \) is a vector of
vertically-averaged velocity components in $x$ and $y$ directions.
Denoted as \( g \) is a gravitational acceleration in the negative vertical
$z$ direction, which is averaged out in the derivation of the shallow water equations.

The sole purpose of presenting this new system of equations is to
demonstrate the flexibility
of our SF-PIF scheme, in that the system-free approach
allows an easy code implementation
without the need for another analytical calculation of
new Jacobian and Hessian terms of the new governing system.
By the system-independent property of the SF-PIF method,
the process of changing from the 2D Euler code to
the 2D SWE code is no more than
switching  the governing equations.
This process is much simpler than
what is needed in the original PIF method,
enabling the overall code transition is highly smooth and transparent
without any extra effort to modify a substantial portion of code lines.

We conduct a simple circular dam breaking simulation,
which has been widely adopted for code validation purposes
in many SWE
literatures~\cite{alcrudo1993high,toro2001shock,delis2005numerical}.
Initially, a volume of still water is confined
in the virtual (i.e., invisible) cylindrical
wall with a radius of 11 meters (m), 
located at the center of simulation domain,
\( \left[ \SI{50}{\meter} \times \SI{50}{\meter} \right] \).
The depth of the water inside of the wall is
\( \SI{10}{\meter} \) and \( \SI{1}{\meter} \) outside.
This configuration would be considered as an
SWE version of 2D Riemann problem.
Explicitly, the initial condition is given as,
\begin{equation}\label{eq:swe-init}
    \left(h, u, v \right) = \begin{cases}
        \left(\SI{10}{\meter}, 0, 0 \right) & \text{for } r \le \SI{11}{\meter}, \\
        \left(\SI{1}{\meter}, 0, 0 \right) & \text{for } r > \SI{11}{\meter}, \\
    \end{cases}
\end{equation}
where \( r \) is the distance from the center of domain,
\( r = \sqrt{\left( x - \SI{25}{\meter} \right)^{2} + \left( y - \SI{25}{\meter} \right)^{2} } \).
The outflow boundary condition is applied for both directions,
and we set the gravitational acceleration \( g = \SI{9.81}{\meter\per\square\second}\).
Again, the Courant number is set to \( C_{\text{cfl}} = 0.4 \).

\begin{figure}[ht!]
    \centering
    \includegraphics[width=160mm]{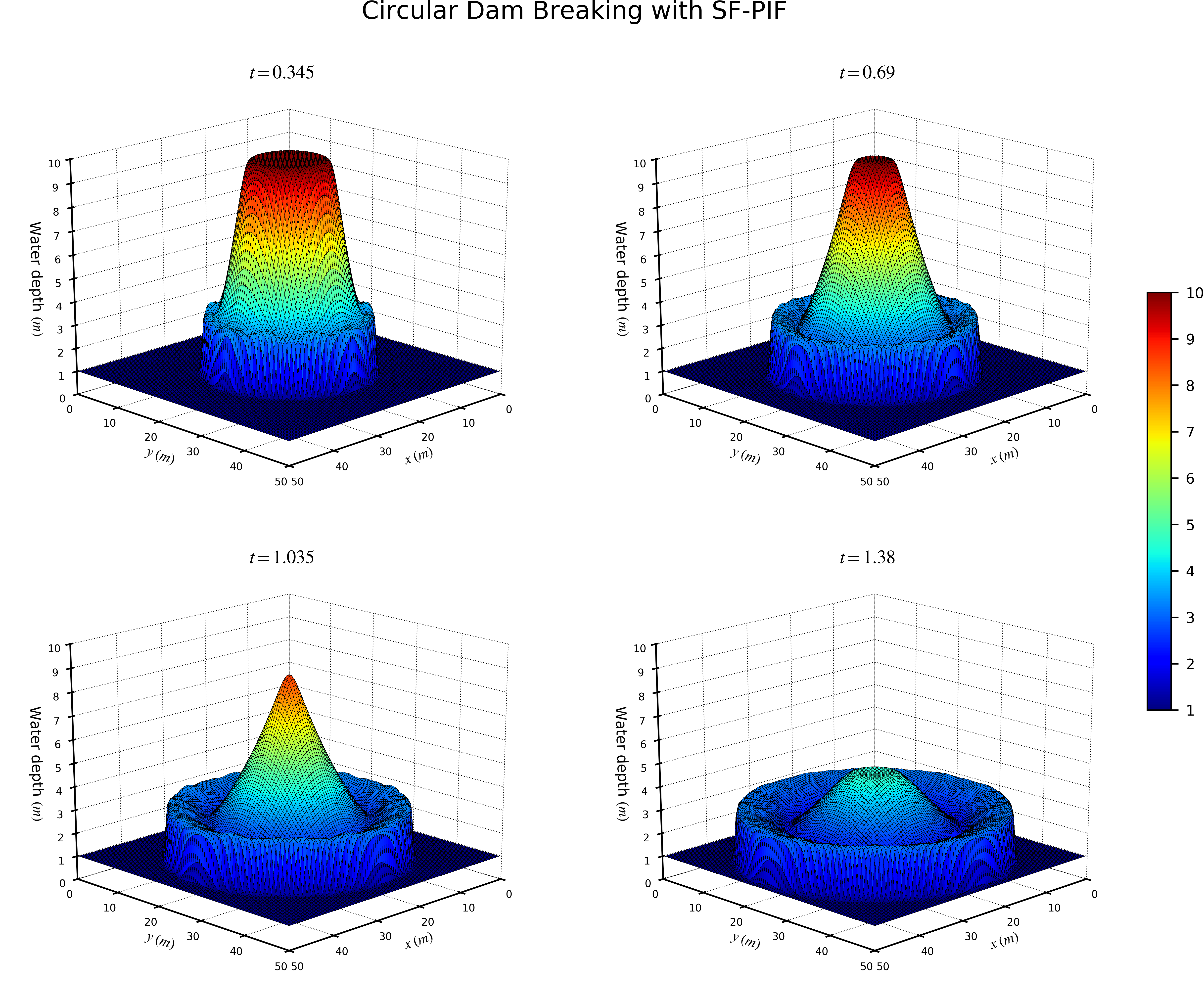}
    \caption{Snapshots of the circular dam breaking simulation
        at four different times, \( t = 0.345, 0.69, 1.035, \) and $1.38$ seconds.
        A volume of water in rest is initially confined
        in a cylindrical dam of a radius $r=11$ m and a height $h=10$ m.
        The simulation starts with an instantaneous removal of the cylindrical wall
        located at the center of the domain
        \( \left[ \SI{50}{\meter} \times \SI{50}{\meter} \right] \).
        The gravitational force is exerted on the steady water,
        which triggers the onset of the gravitational collapse of the entire volume of water,
        making the circular splash in the outer rim as well as the ripple effects
        in the central region that move radially outward in time.
    }\label{fig:swe_circ}
\end{figure}
As shown in Fig.~\ref{fig:swe_circ}, the results with SF-PIF are
well comparable with the results using the same configuration
reported in~\cite{alcrudo1993high,toro2001shock,delis2005numerical}.
We conclude that the overall spherical symmetry and the sharp profile
at the wave front are well maintained in the snapshots
at four different times, $t=0.345, 0.69, 1.035$, and $1.38$.

\section{Conclusion}\label{sec:conclusion}

In this paper, we have improved the flexibility of the
PIF method in~\cite{christlieb2015picard}
by introducing a new Jacobian-free and Hessian-free approach.
With this improvement, referred to as the system-free (SF) approach,
the resulting SF-PIF method
can be readily implemented in any existing RK-based FDM codes
by swapping any multi-stage RK integrators
with the single step, third-order accurate SF-PIF integrator.

The major advantage of SF-PIF lies in ease of its code implementation
for practical use.  In particular, SF-PIF can be applied easily
to a different hyperbolic system of equations without hassle
by virtue of our system-independent formulation of
Jacobian-vector and Hessian-vector-vector multiplications,
which operationally replaces the analytical expressions
of the Jacobians and Hessians terms
in the original PIF method.

In the present study, we have tested our SF-PIF algorithm
to solve a wide range of well-known benchmark test problems in
one and two spatial dimensions. The test results show that
the solution accuracy and stability of SF-PIF are proved to be
equally comparable with the solutions computed using arguably the
most popular choice of the three-stage SSP-RK3 solver.
Moreover, compared to RK3, SF-PIF achieves a twice fast
code performance gain while maintaining the same third-order accuracy.
We also have demonstrated that the SF-PIF method
delivers the same order of accuracy as the original PIF method,
while simplifying the calculations of the Jacobian and Hessian terms.
The two methods are found to be nigh indistinguishable in their solution
accuracy, stability, and performance.
Mathematically speaking, the fact they are identical
assures that the system-free approximation in SF-PIF is
highly reliable in the Jacobian and Hessian term calculations
in comparison with the analytical calculation in PIF\@.
Practically speaking, as the SF-PIF
method does not sensitively rely on a system of governing equations,
it provides an increased adaptability
to effortlessly switch to a different system of equations.
This adaptability has been fully demonstrated
in terms of switching the existing Euler equations code
to a code for the shallow water equations
in Section~\ref{subsec:2d-shallow}.

An extension of the current work 
is to design a fourth-order (or higher)
SF-PIF method. However, a direct extension
to a fourth-order scheme requires complex
vector calculations and evaluations of the next high derivative terms,
higher than the Hessian term in the Taylor expansion.
Such a naive extension will be inevitably facing an increasing complexity
in code implementations, and hence resulting the code performance less attractive
in comparison with the use of conventional RK-based solvers.
Reducing such complexities in designing a fourth- or higher-order
SF-PIF method is essential, and a relevant work
will be further investigated in our future studies.

\section{Acknowledgements}
We acknowledge use of the lux supercomputer at
UC Santa Cruz, funded by NSF MRI grant AST 1828315.
The authors thank Dr. Adam C. Reyes
for very helpful discussions and feedback
during the early stage of the development.

\bibliography{refs}\label{sec:references}

\end{document}